\documentclass[preprint,a4paper,showabstract,superscriptaddress]{revtex4-2}
\setcitestyle{super}

\usepackage{multirow}
\usepackage{graphicx}
\usepackage{epstopdf}
\usepackage{dcolumn}
\usepackage{bm}
\usepackage{hyperref}
\usepackage[utf8]{inputenc} 
\usepackage[T1]{fontenc}    
\usepackage{lmodern}
\usepackage{url}            
\usepackage{booktabs}       
\usepackage{amsfonts}       
\usepackage{nicefrac}       
\usepackage{microtype}      
\usepackage[version=3]{mhchem}
\usepackage{natbib}
\usepackage{amsmath}
\usepackage{amssymb}
\usepackage{xcolor}
\usepackage[range-units = single,range-phrase = \text{--},separate-uncertainty = true,detect-all]{siunitx}
	\DeclareSIUnit\linepair{lp}
	\DeclareSIUnit\pixels{px}
\usepackage{soul}
\setstcolor{red}
\usepackage{cancel}
\def\bibsection{\noindent \textbf{ References.}}

\renewcommand\labelenumi{(\roman{enumi})}
\renewcommand\theenumi\labelenumi

\newcommand*{\short}[1]{\textcolor{black}{#1}}

\newcommand*{\corr}[1]{\textcolor{black}{#1}}
\newcommand*{\revv}[1]{\textcolor{black}{#1}}

\newcommand*{\alex}[1]{\textcolor{black}{#1}}
\newcommand*{\rev}[1]{\textcolor{black}{#1}}
\newcommand*{\ulysse}[1]{\textcolor{black}{#1}}
\renewcommand{\vec}[1]{\bm{\mathrm{#1}}}

\newcommand\out{_\mathrm{out}}
\newcommand\rin{\ulysse{\vec{\rho}_\mathrm{in}}}
\newcommand\rout{\ulysse{\vec{\rho}\out}}

\newcommand\uout{\vec{u}\out}

\begin{document}

\title{Detection and characterization of targets in complex media using fingerprint matrices}

\author{Arthur Le Ber}
\thanks{These authors equally contributed to this work}
\affiliation{Institut Langevin, ESPCI Paris, PSL University, CNRS, Paris, France}
\author{Antton Go\"{i}coechea}
\thanks{These authors equally contributed to this work}
\affiliation{Institut Langevin, ESPCI Paris, PSL University, CNRS, Paris, France}
\author{Lukas M. Rachbauer}
\affiliation{Institute for Theoretical Physics, Vienna University of Technology (TU Wien), Vienna, Austria}
\author{William Lambert}
\affiliation{Supersonic Imagine, Aix-en-Provence, France}
\author{Xiaoping Jia}
\affiliation{Institut Langevin, ESPCI Paris, PSL University, CNRS, Paris, France}
\author{Mathias Fink}
\affiliation{Institut Langevin, ESPCI Paris, PSL University, CNRS, Paris, France}
\author{Arnaud Tourin}
\affiliation{Institut Langevin, ESPCI Paris, PSL University, CNRS, Paris, France}
\author{Stefan Rotter}
\affiliation{Institute for Theoretical Physics, Vienna University of Technology (TU Wien), Vienna, Austria}
\author{Alexandre Aubry}
\thanks{Corresponding author: alexandre.aubry@espci.fr}
\affiliation{Institut Langevin, ESPCI Paris, PSL University, CNRS, Paris, France}

\date{\today}
\begin{abstract}
    \textbf{\short{As waves propagate through a complex medium, they undergo multiple scattering events. This phenomenon is detrimental to imaging, as it causes full blurring of the image. Here, we report a method to detect, localize and characterize any scattering target embedded in a complex medium.  We introduce a fingerprint operator that contains the specific signature of the target with respect to its environment. When applied to the recorded reflection matrix, it provides a likelihood index of the target state. This state can be the target's position for localization purposes, its shape for characterization or any other parameter that influences the target response. We demonstrate the versatility of our method by performing proof-of-concept ultrasound experiments on elastic spheres buried inside a strongly scattering granular suspension and on lesion markers, commonly used to monitor breast tumours, embedded in a foam mimicking soft tissue. Furthermore, we show how the fingerprint operator can be leveraged for characterizing the complex medium itself by mapping the fiber architecture within muscle tissue. Our method is broadly applicable to different types of waves beyond ultrasound for which multi-element technology allows a reflection matrix to be measured.}}
\end{abstract}
\maketitle
\newpage

\noindent {\large \textbf{Introduction}} 

With the emergence of multi-element technology, {it has become possible to cope with the enormous complexity of disordered media for focusing} waves through them or {for imaging} objects hidden behind them. This feat has been demonstrated in acoustics through time-reversal mirrors~\cite{Fink1997} and in optics using wavefront shaping techniques~\cite{mosk_controlling_2012}. A matrix formalism has become particularly useful in this context, especially when wave-fields are controlled by arrays of independent elements \short{both in transmission~\cite{Foschini1998, Popoff2010} and reflection~\cite{Prada1994,Popoff2011}.} 
Since an inhomogeneous medium can be treated as one realization of a random process, some aspects of random matrix theory and basic concepts {from scattering theory} have been fruitfully applied to wave control or imaging through complex media~\cite{rotter_light_2017,Cao2022,bertolottiImagingComplexMedia2022}. In {particular}, transmission eigenchannels have been shown to provide a properly designed combination of incident waves that can be fully transmitted through a disordered medium~\cite{Gerardin2014, popoffCoherentControlTotal2014,Davy2015}. 
{The reflection eigenchannels, on the other hand, have been shown to focus on bright point-like targets embedded in a scattering medium, provided that the disorder is not too strong~\cite{Aubry2009,Badon2016}. For more complex targets, these \short{approaches} are not adapted since the one-to-one association between each eigenstate and each target is no longer verified in the general case~\cite{Aubry2006,Robert2009}. Much progress has also been made using \short{the reflection matrix}~\cite{Yoon2020,joThroughskullBrainImaging2022,leeHighthroughputVolumetricAdaptive2022,Weinberg2024,Badon2020,Lambert2020} for imaging\alex{, in particular, to compensate the effects of disorder~\alex{\cite{Bureau2023,Murray2023,Giraudat2024,zhangDeepImagingScattering2023,Najar2024}}}. All currently available techniques, however, face the key challenge that, in the regime of strong multiple scattering, such a compensation becomes close to impossible.}


In the present {article}, {we address this outstanding challenge with an approach that is not based on compensating the effect of disorder, but rather by detecting correlations in the scattering of waves that survive even for very thick disordered media. The starting point for this approach is the following scattering invariant operator that has been constructed based on the matrix product $\mathbf{T} \times \mathbf{T}_0^{\dag}$ between the transmission matrix of a disordered medium $\mathbf{T}$ and a matrix $\mathbf{T}_0$ of a homogeneous reference medium~\cite{Pai2021}. The associated eigenstates of this operator have been shown to be `scattering invariant modes' in the sense that they exhibit a transmitted field pattern behind a disordered medium that is identical to that of purely ballistic waves, independent of the multiple scattering events endured by the waves inside the medium.}

\short{In the context of target detection,} we are looking for input states that are reflected from a target embedded inside a disordered medium in the same way as from a target embedded in free (or homogeneous) space. Correspondingly, the operator \alex{$\mathbf{\Gamma}$} that we introduce here is defined as the matrix product $\alex{\mathbf{\Gamma}} = \mathbf{R} \times \mathbf{R}_{0}^{\dagger}$ between a measured reflection matrix $\mathbf{R}$ of \alex{a disordered medium potentially hiding a target} and the conjugate transpose of a reference matrix \alex{$\mathbf{R}_0$} \alex{associated with a target in absence, this time, of any disordered environment}. This \short{fingerprint} matrix \alex{$\mathbf{R}_0$} \alex{can be parametrized by a target state vector $\mathbf{q}$} whose coefficients account for the position, size and shape of the target alone.
\short{The} more closely the position, size or shape of the target in the fingerprint matrix $\mathbf{R}_0(\mathbf{q})$ match their true values realized in the experimentally measured matrix $\mathbf{R}$, the higher the correlations between these two matrices will be - allowing us to determine the target properties through an optimization problem.
{Our} concept is therefore extremely versatile since the {fingerprint matrix} can be designed as a function of the experimental situation, and can be applied to any kind of waves, provided that a measurement of the reflection matrix is possible. 

\short{To demonstrate the capabilities and versatility of the fingerprint operator, we conduct three ultrasound imaging experiments. First, we detect and localize spherical metal targets buried within a highly scattering granular medium. Second, we demonstrate the detection of a clinical lesion marker embedded in a scattering tissue-mimicking phantom. Finally, we show that the fingerprint operator enables the extraction of structural information such as orientation and size of anisotropic features in biological tissue, exemplified by the in vivo imaging of muscle fiber architecture. These three examples highlight the robustness and flexibility of the approach under various scattering conditions and target complexities.}
\vspace{5mm}

\noindent {\large \textbf{Results}} 

\noindent {\textbf{Multiple scattering as a nightmare for imaging}} 

\begin{figure*}[ht]
\centering
\includegraphics[width=\textwidth]{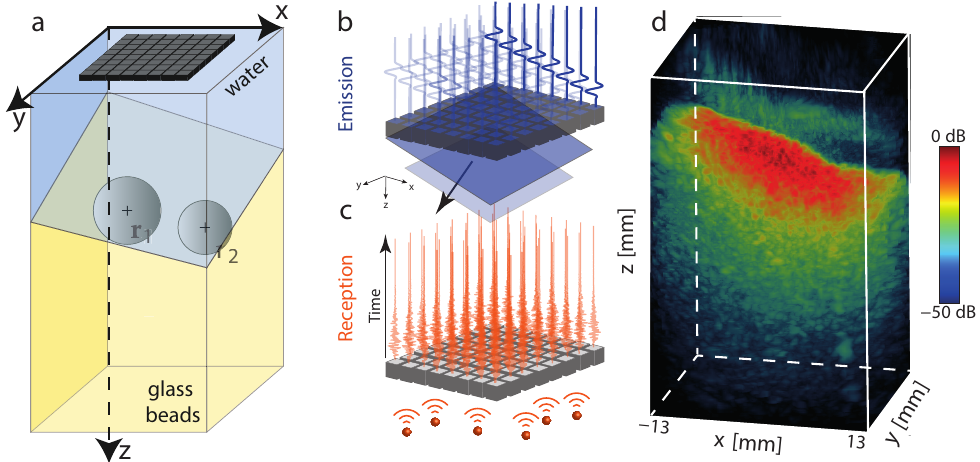}
\caption{\label{config} \textbf{Multiple scattering fog.} \textbf{a}, Experimental configuration{: A 2D array of 1024 ultrasound transducers {acting both as transmitters and receivers} is used to detect two metal spheres that are embedded in strongly scattering glass beads below water.} \textbf{b}, The reflection matrix is measured {by insonifying the medium with a set of incident plane waves ($\bm{\theta}_{\textrm{in}}$)}. \textbf{c}, Each back-scattered wave field $R(\mathbf{u}_{\textrm{out}},\bm{\theta}_{\textrm{in}},t)$ is recorded on each transducer $\mathbf{u}_{\textrm{out}}$ of the {same probe}. \textbf{d}, Volumetric visualization of the probed medium by projecting the intensity maximum of the confocal image obtained by a delay-and-sum beamforming applied to the recorded reflection matrix (a dynamic view of this is shown in Supplementary Movie 1). {With this conventional imaging technique, {the specular echo of the interface between water and the glass bead suspension largely predominates and the} targets are completely invisible on the image.}}
\end{figure*}

The first experiment consists in the ultrasound imaging of two {metal spheres} embedded into a granular suspension {consisting of} randomly packed glass beads (diameter $\varnothing \simeq$ 300-315 $\mu$m) immersed in water {with a packing density of 60\%}~\citep{Wildenberg2019} (Fig.~\ref{config}a). The {{metal spheres} have} diameters $d_1=10$ mm and $d_2=8$ mm. Their centers $\mathbf{r}_1$ and $\mathbf{r}_2$ are positioned 9 and 7 mm below this granular {suspension} surface, respectively. The experimental procedure first consists in the acquisition of the {reflection} matrix {using} a {two}-dimensional array of 1024 transducers placed on top of the {granular suspension} surface into which the {metal} spheres are embedded (Methods). The reflection matrix is here acquired using a set of plane waves~\cite{Montaldo2009} (Fig.~\ref{config}b, Methods) over a 1.8-2.6 MHz frequency bandwidth \short{(Extended Data Tab.~\ref{ProbeInfo})}. For each plane wave {with incidence angles} $\bm{\theta}_{\textrm{in}}=(\theta_x,\theta_y)$, the time-dependent reflected wave field  $R(\uout,\bm{\theta}_{\textrm{in}},t)$ is recorded by each transducer $\uout$ (Fig.~\ref{config}c). This set of wave-fields forms a reflection matrix acquired in the plane wave basis, $\mathbf{R}_{\bm{\theta}\mathbf{u}} (t)=\left [ R(\uout,\bm{\theta}_{\textrm{in}},t) \right ]$. A Fourier transform can then be applied to the experimentally acquired reflection matrix to obtain its frequency-dependent counterpart, $\mathbf{R}_{\bm{\theta}\mathbf{u}} (f)=\int dt \mathbf{R}_{\bm{\theta}\mathbf{u}} (t) \exp\left (-j2\pi f t\right) $, with $f$ the frequency.  

A focused beamforming algorithm (Methods) is then applied to the reflection matrix to obtain a confocal image of the medium under investigation. The result is displayed in Fig.~\ref{config}d. Although the image shows a strong specular echo at the surface of the granular {suspension}, the two {spheres located} beneath this surface are fully invisible on the image. Two reasons can be invoked to explain the failure of a standard imaging process in this experimental configuration:
\begin{enumerate}
    \item First, the granular {suspension} is strongly scattering, as evidenced by the scattering mean free path $\ell_s \sim 1.8$ mm exhibited by ultrasonic waves in the frequency range of interest (Supplementary Section~S1). Note that scattering by the glass {bead packing} is isotropic in the frequency range under study. The transport mean free path is therefore equivalent to $\ell_s$. The imaging depth $z$ of each {sphere} ranges from 4 to 5$\ell_s$. At such depths, the direct echoes of the {target} {sphere}s are hidden by a strongly predominant multiple scattering {background}, which generates an intense speckle fog in Fig.~\ref{config}d. The image contrast $\mathcal{C}_I$, which scales as the single scattering rate, undergoes a drastic exponential decay with the penetration depth $z$ ~\cite{Goicoechea2024}:
\begin{equation}
\label{ball}
\mathcal{C}_I \propto \exp \left (-z/\ell_s \right )
\end{equation}
The target echo is thus decreased by a factor ranging from $50$ ($z\sim$4$\ell_s$) to $150$ ($z\sim$5$\ell_s$) with respect to the multiple scattering background in the present experiment.
    \item Second, a confocal imaging scheme is far from being optimal for the imaging of objects such as the {elastic} {sphere}s considered here. Each of them actually supports a large number of resonating modes {or/and internal waves} giving rise to a complex spatio-temporal echo. 
\end{enumerate}  
\vspace{2 mm}

\short{\noindent { \textbf{\short{Leveraging correlations through} the fingerprint operator}} }

{Even if the} response of {an elastic target} may be spectrally and spatially quite complex~\cite{Thomas1994,Prada1998,Aubry2006,Robert2009,Gespa1987}, all these features can be adequately captured by measuring the target's reflection matrix \short{$\mathbf{R}_0$ in free space (Methods, Extended Data Fig.~\ref{free})} {and leveraged by the \short{scattering-invariant} operator \short{$\mathbf{\Gamma} = \mathbf{R} \times \alex{\mathbf{R}_0}^{\dagger}$}}. 

Rather than inspecting individual eigenstates and eigenvalues of $\mathbf{\Gamma}$, however, we quantify the correlation signatures through the trace of this matrix, integrated over {the} entire recorded frequency interval and properly normalized:
\begin{equation}
\alex{\gamma}(\mathbf{q})= \frac{\left| \int df \textrm{Tr} \left \lbrace {\mathbf{R}}(f) \times \alex{\mathbf{R}_0}^{\dag}(\mathbf{q},f)  \right \rbrace \right|}{\left [ \int df  ||{\mathbf{R}}(f)  ||_F \times  \int df ||\alex{\mathbf{R}_0}(\mathbf{q},f)  ||_F \right ]^{1/2} },
\label{eq P}
\end{equation}
where the symbol $||\cdot ||_F$ stands for the {Frobenius} norm.
{The likelihood index $\alex{\gamma}(\mathbf{q})$ for the target to be in state $\mathbf{q}$, involves an emulated target fingerprint matrix \alex{$\mathbf{R}_0(\mathbf{q})$}, where $\mathbf{q}$ accounts for the target state such as, for instance, the target's position $\mathbf{r}$, size, shape or any other parameter that has an influence on the target's response.}

{The question relevant to the practical implementation of this approach} is now the construction of {the} fingerprint operator. The most direct {way of doing it} is to consider the free space reflection matrix {$\mathbf{R}_0(\mathbf{r}_0)$} shown in Extended Data Fig.~\ref{free} and {to} emulate a {fingerprint matrix} $\mathbf{R}_0(\alex{\mathbf{r}})$ by a virtual shift of the target from the initial position $\mathbf{r}_0$ to any point $\mathbf{r}$ through simple matrix operations (Methods). A first validation of Eq.~\ref{eq P} can be performed by applying it to the free-space reflection matrix such that $\mathbf{R}=\mathbf{R}_0\alex{(\mathbf{r}_0)}$. The result is displayed in Extended Data Fig.~\ref{free}f. The comparison with the confocal image displayed in Extended Data Fig.~\ref{free}e highlights the benefit of the proposed approach. All the energy {radiated by the elastic} {sphere} that was initially dispersed in time and space on the diagonal and off-diagonal coefficients of the focused $\mathbf{R}-$matrix is now concentrated onto a single pixel of the image.

The gain $G$ in contrast compared to the confocal image is drastic since it scales as the product between the numbers of spatial and temporal d.o.f, $N_S$ and $N_T$, exhibited by the target (Supplementary Section~S4):
\begin{equation}
\label{G}
G \sim N_S \times N_T
\end{equation}
On the one hand, $N_T$ can be expressed as the product of the frequency bandwidth $\Delta f$ and the reverberation time $\Delta t$ of the target echo:
\begin{equation}
N_T \sim \Delta f \Delta t.
\end{equation}
In the present case, the reverberation time $\Delta t$ lasts 40 $\mu$s for each {sphere} which yields $N_t\sim 32$. On the other hand, $N_S$ is the effective rank of the target reflection matrix $\mathbf{R}_0$, that can be assessed from the singular value distribution of $\mathbf{R}_0$ (Eq.~\ref{svd}, Supplementary Figure~S8): $N_S\sim 20$ for {sphere} 1 and $N_S\sim 12$ for {sphere} 2. This larger value for {sphere} 1 is explained by an effective rank of $\mathbf{R}_0$ scaling as the number of lateral resolution cells covered by the target~\cite{Aubry2006,Robert2009}:
\begin{equation}
N_S \sim \mathcal{A}/ \delta \rho_D^2
\end{equation}
with $\mathcal{A}$, the target physical cross-section and $\delta \rho_D \sim \lambda/(2 \sin \alpha)$, the diffraction-limited resolution length of the target image in Extended Data Fig.~\ref{free}, and $\alpha$ the angle under which the {target} is seen by the array. The resolution length actually determines the transverse size of the focal spot displayed by the $\gamma-$map in Extended Data Fig.~\ref{free}. 

In the present case, given the values of $N_S$ and $N_T$ provided above, the fingerprint operator can, in principle, increase the {sphere} contrast by a factor $G$ (Eq.~\ref{G}) ranging from 380 ({sphere} 2) to 640 ({sphere} 1). {This contrast enhancement is significantly larger than the decay factor experienced by each sphere echo with respect to multiple scattering: 50 for sphere 1 and 150 for sphere 2 (Eq.~\ref{ball})}. {The fingerprint operator can therefore be expected to detect and locate the spheres in the experiment shown in Fig.~\ref{config}.}\\ 

\noindent { \textbf{Seeing into the scattering fog}} 

\begin{figure*}[ht]
\centering
\includegraphics[width=\textwidth]{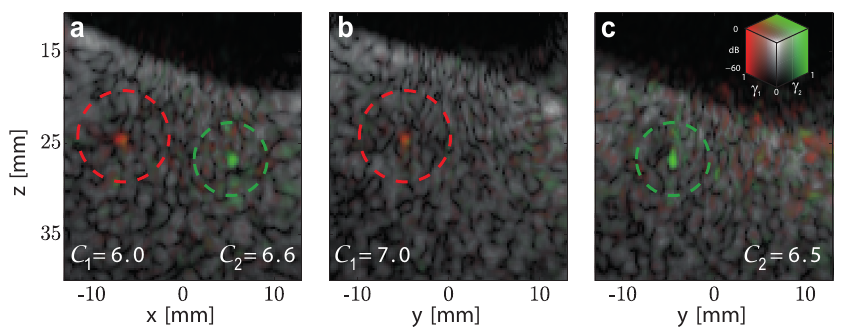}
\caption{\label{result} \textbf{Detecting and localizing target {metal spheres} hidden in the multiple scattering fog.} Different cross-sections of the $\gamma$-map for each {sphere} {(diameter $d_1 = 10$ mm in red, $d_2 = 8$ mm in green)} are superimposed to the corresponding confocal image (B\&{W} scale in dB). \textbf{a} $(x,z)-$cross-section at $y=y_1=y_2$ (\short{scale bar: 10 mm)}. \textbf{b} $(y,z)-$cross-section at $x=x_1$. \textbf{c} $(y,z)-$cross-section at $x=x_2$. {The confocal images correspond to cross-sections of Fig.~\ref{config}d, the dashed circles correspond to the known radius of each sphere. The value of the contrast \alex{$\mathcal{C}_\gamma$} at the target's position with respect to the average value of \alex{$\gamma$} outside {each target} is also indicated. While the targets are completely invisible in the confocal images, they appear very well localized through the calculation of their associated likelihood index $\gamma$. }}
\end{figure*}

{As we {will show in the following}, the target's signatures can be made even more specific with respect to its environment by \alex{considering a generalized scattering invariant operator, $\hat{\mathbf{\Gamma}}=\mathbf{R}\times \mathbf{F}^{\dag}$, involving a more selective fingerprint matrix ${\mathbf{F}}$}  (Supplementary Section~S2).} To that aim, different filters are applied to the {reference} matrix $\mathbf{R}_0$ to build the {matrix} $\mathbf{F}$ (Methods) \short{and make the response of the target more specific with respect to its environment.}
This optimized {fingerprint} operator $\mathbf{F}$ is then used to compute the likelihood maps of the 8 mm- and 10 mm-diameter {sphere}s in the experimental configuration. {To highlight the drastic improvement provided by our approach, we superimpose the corresponding} \alex{$\gamma$}-maps encoded in green and red, respectively, onto the confocal image in Fig.~\ref{result}{, where} different cross-sections of the imaged volume are displayed{:} While the confocal image is fully blurred by the multiple scattering fog, each likelihood map allows the unambiguous detection of each {sphere}. 

{The scattering invariant operator tends to filter out the diffuse waves and select the paths that are insensitive to scattering, \textit{i.e.} paths that behave as if we were in free space~\cite{Pai2021} (among which the coherent ballistic paths). Even though these paths correspond to an extremely weak contribution to the total wave-field in the diffusive regime, the scattering-invariant operator allows us to leverage them for target detection. This is made possible by the complexity of the target signature: The weight of scattering-invariant paths is enhanced by the number of spatio-temporal degrees-of-freedom exhibited by the target echo (Eq.~\ref{G}). {Here, the strong contrast enhancement provided by the fingerprint operator partially relies on the strong impedance mismatch between the metal spheres and the granular suspension. Weaker enhancements are expected for softer materials.}}

{In Supplementary Movie 2, we show how the scattering invariant operator can also work in a dynamic scenario where the target (sphere 1) sinks inside the vibrated dense granular suspension that then acts as quicksand. This experiment was already described in a previous work~\cite{Wildenberg2019} for a granular suspension of smaller glass beads (100 $\mu$m-diameter) that was almost transparent for ultrasound. Here, thanks to the fingerprint matrix concept, we are able to track the trajectory of the sphere in a much stronger scattering regime.} \\

\noindent {\textbf{Detection rate and localization precision}}

{Figure~\ref{result} demonstrates that} the exponential decrease of the target echo {(captured by $\mathcal{C}_I$ in} Eq.~\ref{ball}) is largely compensated by the gain $G$ (Eq.~\ref{G}) provided by the fingerprint operator. Experimentally, the target contrast \alex{$\mathcal{C}_\gamma$} can be evaluated by considering the ratio between the value of $\gamma$ at each target position and its average outside of each target. A significant contrast ranging between 6 and 7 is found for each {metal} {sphere} (see $\mathcal{C}_\gamma$ values indicated for each cross-section {in} Fig.~\ref{result}). Those values are in qualitative agreement with the theoretical prediction: $\mathcal{C}_\gamma\sim G \times \mathcal{C}_I\sim$ 4.2 for sphere 1 and 7.6 for sphere 2. 

{The observed contrast is an important feature since it allows to compute a probability of false alarm $PFA$, \textit{i.e.} the probability that the bright spots in the $\gamma$-maps of Fig.~\ref{result} correspond to a normal multiple scattering fluctuation and not to coherent targets. Assuming Rayleigh statistics for multiple scattering speckle, the observed contrast values ($\mathcal{C}_{\gamma}\sim 6-7$) indicate an extremely small probability of false alarm: $PFA < 10^{-8}$ (see Supplementary Section~S6 and Figure~S15). Hence, the detection of metal spheres is achieved with an extremely high degree of confidence in Fig.~\ref{result}.}

{Besides providing a detection rate, the contrast $\mathcal{C}_{\gamma}$ also dictates} the precision $\delta \rho_L$ of the localization process predicted by the Cram\'{e}r-Rao bound~\cite{Quazi1981,Desailly2015} (Supplementary Section~S5): 
\begin{equation}
\label{loc}
 {   \delta \rho_L \sim \frac{\delta \rho_D}{ \sqrt{\mathcal{C}_\gamma}} \sim \frac{\delta \rho_D}{\sqrt{G\mathcal{C}_I}}}.
\end{equation}
{By increasing the target contrast with respect to its environment, the scattering invariant operator enhances the localization precision by a factor scaling as \alex{$\sqrt{N_T N_S}$ (Eq.~\ref{G})}. The more complex the target \alex{echo is}, the more beneficial our method becomes.}

{This contrast-based evaluation of detection and localization performance can also be used to evaluate the robustness of the method with respect to electronic noise that would unavoidably pollute the recorded data in presence of absorption. To do so, the initial contrast $\mathcal{C}_{I}$ can be replaced by a signal-to-noise ratio (SNR). The detection threshold is then inversely proportional to the factor $G$ (Eq.~\ref{G}) and the localization precision (Eq.~\ref{loc}) scales as $\delta \rho_L \sim {\delta \rho_D}/{ \sqrt{G \times \textrm{SNR}}}$.}\\

\noindent {\textbf{Lesion marker detection and localization in ultrasound speckle}}

\begin{figure*}[ht]
\centering
\includegraphics[width=12cm]{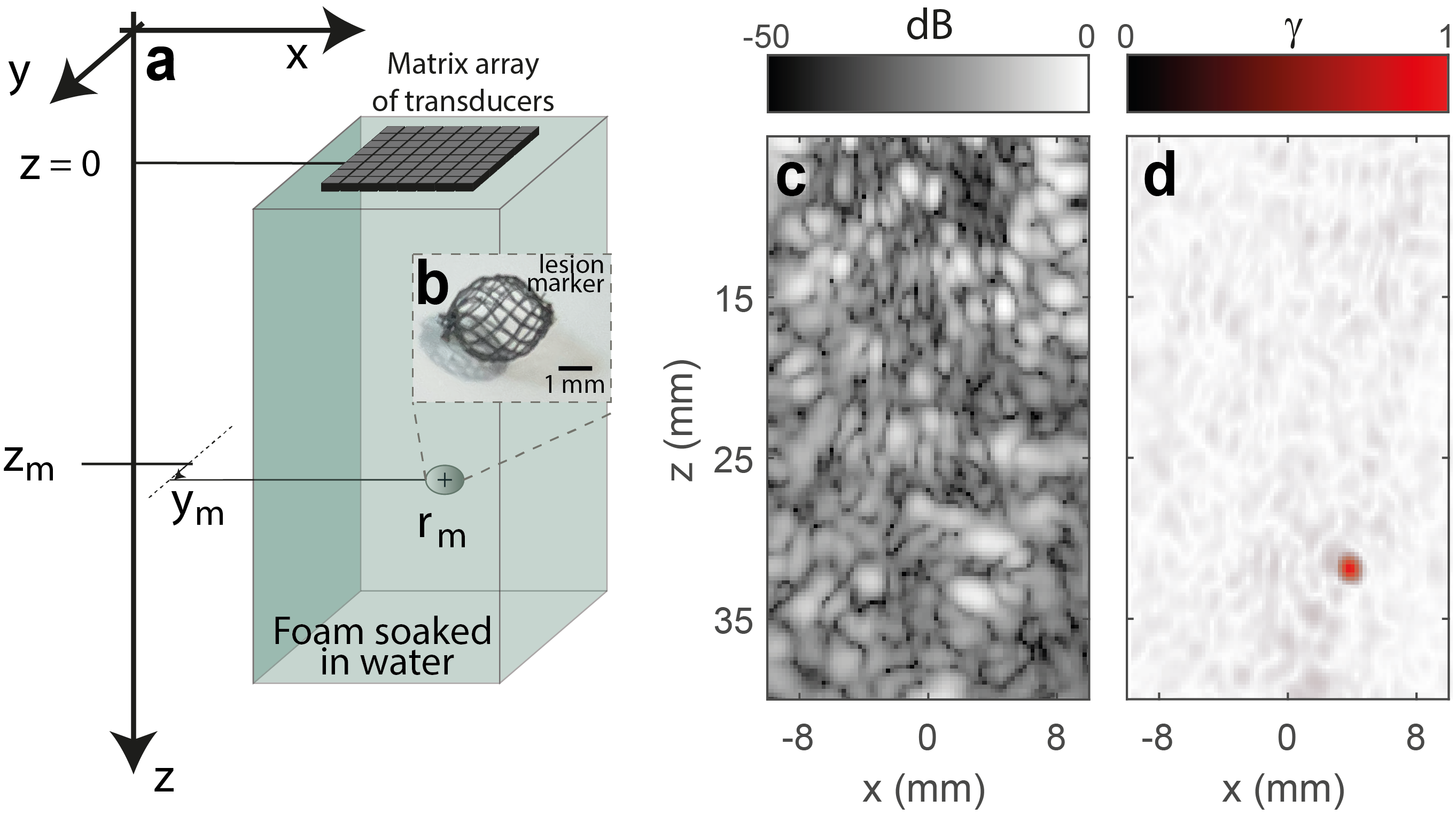}
\caption{\label{lesion} \textbf{Localizing a lesion marker in ultrasound speckle.} \textbf{a}, Experimental configuration: The $32\times 32$ probe is used to image a lesion marker embedded into  foam soaked in water. The position $\mathbf{r}_m=(x_m,y_m,z_m)$ of the lesion marker center is $(3.5,4.5,32)$ mm.  \textbf{b}, Photography of the lesion marker (credit: Arthur Le Ber). \textbf{c}, Longitudinal cross-section of the {confocal} ultrasound image in the plane $y=y_m$ \short{(scale bar: 5 mm)}. \alex{\textbf{d}, Corresponding likelihood map of the lesion marker superimposed {on}to the confocal image in transparency.}}
\end{figure*}

To show the generality of the concept and its potential applications, a second experiment depicted in Fig.~\ref{lesion}a (Methods,  \short{Extended Data Tab.~\ref{ProbeInfo}}) was carried out {in a weaker scattering regime ($z\sim \ell_s$) characteristic of ultrasound imaging and in which single and multiple scattering coexist~\cite{Goicoechea2024}}. It involves the detection of a lesion marker (Tumark\textregistered Vision, Somatex, Fig.~\ref{lesion}b) generally used to monitor breast tumors in {clinical settings}~\cite{Rueland2018}. Although {these markers have} been specifically designed to yield a clear signature on ultrasound images, {their} detection and localization are often hampered by the ultrasound speckle generated by unresolved scatterers {that are randomly distributed} in tissue. In the experiment depicted in Fig.~\ref{lesion}, ultrasound speckle is generated by a foam soaked in water in which the lesion marker has been embedded. \alex{This foam generates an ultrasound speckle similar to {that of} soft tissues.} \alex{The corresponding confocal image highlights this characteristic speckle which here prevents an unambiguous detection and localization of the target (Fig.~\ref{lesion}c)}. On the contrary, the fingerprint operator designed for this lesion marker allows a clear detection and sharp localization of the lesion marker \alex{(Fig.~\ref{lesion}d)}. This proof-of-concept shows the potential of the fingerprint operator for the monitoring of any object (\textit{e.g.} needle, catheter, marker, etc.) used in interventional radiology.\\

\noindent { \textbf{Towards quantitative imaging of complex media}} 

Beyond the detection and localization of a {known} intruder inside a complex medium, the fingerprint operator can {also} be {applied to unknown targets. In that case, a dictionary of fingerprint matrices can be analytically or numerically built for targets of arbitrary shape, size, orientation \textit{etc.} A maximization of the likelihood index with respect to these parameters can then lead to a quantitative characterization of the inspected medium.}

\begin{figure*}[ht]
\centering
\includegraphics[width=\textwidth]{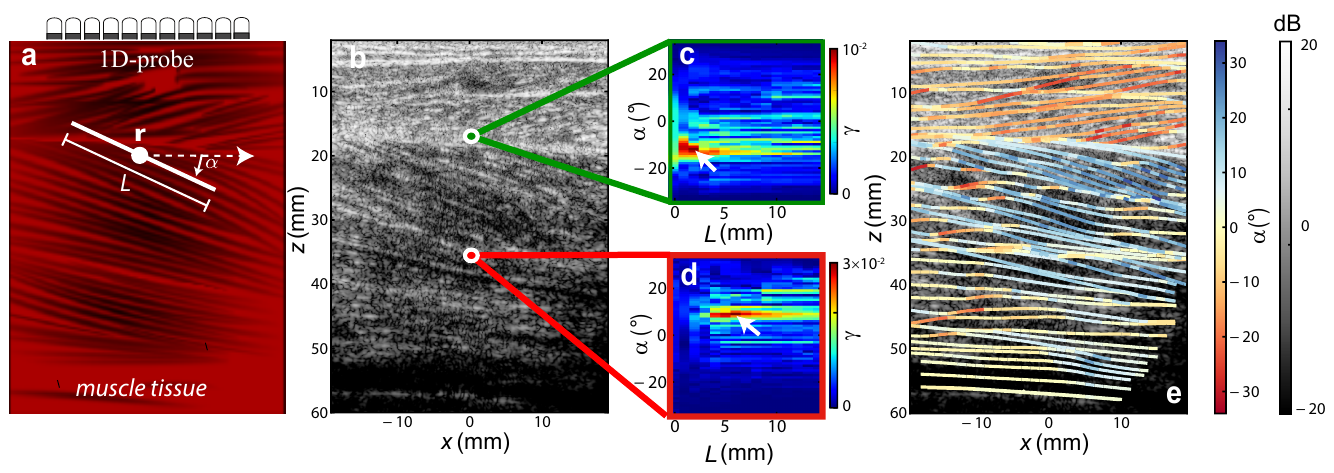}
\caption{\label{muscle} \textbf{Revealing the local architecture of fibers in muscle tissue.} \textbf{a}, {Sketch of the} experimental configuration: An array of transducers is placed in contact with the calf of a healthy patient. The fingerprint operator \alex{$\mathbf{R}_0(\mathbf{q})$} here corresponds to a set of reflection matrices associated with a mirror of size $L$ and orientation $\alpha$ simulated for each pixel $\mathbf{r}$ in the medium considered as homogeneous  ($c_0 =$ 1580 m/s). \textbf{b}, Ultrasound image of the calf in dB. {\textbf{c}, \textbf{d}, Likelihood index $\gamma$ in the $(L, \alpha)$ parameter space for two positions yielding different values for the \alex{local correlation} length {$L$} and fiber orientation {$\alpha$}. The white arrow{s} indicate the optimal length $L_{\textrm{opt}}$ and orientation $\alpha_{\textrm{opt}}$ at each position.} \textbf{e}, Vector representation of the fibers deduced from the orientation distribution $\alpha_{\textrm{opt}}(\mathbf{r})$ encoded with the color scale displayed on the right.}
\end{figure*}

{As a proof-of-concept,} we will show how it can be used for mapping the local anistropy of a fibrous medium. This information is particularly relevant for muscle tissues in ultrasound imaging in order to diagnose neuromuscular~\cite{Wijntjes2020} or myocardial~\cite{Papadacci2017} diseases. 
The experiment is carried out on a {human calf, in-vivo,} the fibers of which are partially visible on the ultrasound image displayed in Fig.~\ref{muscle}b. This image is built from the reflection matrix $\mathbf{R}$ recorded using a linear phased array of 256 transducers over the [5.5; 9.5] MHz frequency bandwidth (Methods, \short{Extended Data Tab.~\ref{ProbeInfo}}). 
In order to image fibrous tissues, the fingerprint {matrix} is defined as a free-space reflection matrix $\mathbf{R}_0(\mathbf{q})$ associated with a reflecting {1D} mirror whose state $\mathbf{q}$ is described by its position $\mathbf{r}$, its orientation $\alpha$ and its dimension $L$ (Fig.~\ref{muscle}a). The dictionary of fingerprint matrices $\mathbf{R}_0(\mathbf{q})$ is constructed numerically (Methods).

The \alex{scalar} {product} of the recorded reflection matrix $\mathbf{R}$ {with} the fingerprint {matrix} \alex{$\mathbf{R}_0$} \alex{(Eq.~\ref{eq P})}  provides a likelihood index $\alex{\gamma}(\mathbf{r},\alpha,L)$ with respect to the parameters $\alpha$ and $L$ at each point $\mathbf{r}$ in the image (Eq.~\ref{eq P}).
The dependence of the $\alex{\gamma}$-map with respect to parameters $\alpha$ and $L$ is shown for two points $\mathbf{r}$ in the field-of-view in Figs.~\ref{muscle}c and d. The maximum value of this quantity gives the orientation $\alpha_{\textrm{opt}}(\mathbf{r})$ and the local \alex{correlation} length $L_{\textrm{opt}}(\mathbf{r})$ of the fibrous medium at each point $\mathbf{r}$:
\begin{equation}
\lbrace \alpha_{\textrm{opt}}(\mathbf{r}), L_{\textrm{opt}}(\mathbf{r})
\rbrace =    \arg \max_{\lbrace \alpha,L \rbrace } \left [ \gamma(\mathbf{r},\alpha , L) \right]
\end{equation}
The size and orientation distribution of the fibers provide a vector representation of the fibers superimposed to the confocal image in Fig.~\ref{muscle}e. A very good agreement is found between the visual appearance of the fibers on the ultrasound image and their vector representation provided by the fingerprint operator. 


This information could be extremely rewarding to monitor neuro-muscular diseases~\cite{Wijntjes2020} and myocardial fiber disarray that emerges in the early stage of many pathologies such as in cardiomyopathies or in fibrosis~\cite{Tseng2005}. It can also be relevant in the context of non-destructive testing for determining the orientation of elongated grains in polycrystalline materials~\cite{Thompson2008} or, in optical microscopy, for probing the collagen orientation and its lamellar distribution~\cite{Raoux2023}.\\

\noindent {\large \textbf{Discussion}} \\

In this {manuscript}, we have introduced the fingerprint operator with the aim to detect, localize and characterize any object in a {strongly scattering medium}. A necessary condition, however, is the creation of a sufficiently complete and precise knowledge of the reflection matrix and its dependence according to a state of interest. {From} that perspective, the generalized polarization tensor, that was introduced ten years ago for electromagnetic waves, can be of interest~\cite{Ammari2013,Ammari2014}. 
\short{However}, this approach requires an array of sensors fully surrounding the target under investigation{, which is not available} in most in-vivo or in-situ applications. On the contrary, the fingerprint operator {works in reflection and} can be adapted to any experimental configuration, thereby providing {much higher} robustness with respect to the non-ideal nature of experiments. {Moreover, the simplicity of our approach makes it relatively inexpensive in terms of computational time and power (Methods). Indeed, the construction of a $\gamma-$map is computationally equivalent to a standard beamforming process used to build a confocal image (Supplementary Section~S7).}   

{An important question is the application of the fingerprint operator to dynamic imaging scenarios in which the target motion cannot be neglected during the measurement of the reflection matrix measurement. Again, the flexibility of the fingerprint matrix can help since it can be defined as a function of target motion parameters (motion direction, velocity, acceleration). A maximization of the likelihood index $\gamma$ with respect to these parameters would then provide a quantification of target motion with a contrast enhancement equivalent to a static scenario but at the expense of a larger computational burden (Methods).}

Beyond its versatility, the fingerprint operator is also a universal concept that can find applications in all fields of wave physics where multi-element technology is available. MIMO radar~\cite{Xu2008} and sonar~\cite{Pailhas2017} are examples of fields where target identification and localization in noisy environments is a long-lasting challenge. \short{The fingerprint operator is also extremely promising for characterizing objects whose spatio-temporal response depends on their environment, thereby allowing a local measurement of intensive physical quantities, such as temperature or pressure. Microbubbles~\cite{DeJong2002,TremblayDarveau2014,errico_ultrafast_2015} in ultrasound imaging or plasmonic nanoparticles~\cite{Anker2008} in optics are ideal candidates for this purpose.} 
{While this is by no means an exhaustive list}, the flexibility of the fingerprint operator concept {can be critical to all of these applications}.  \\
\vspace{5mm}


\setcounter{figure}{0}
\renewcommand{\figurename}{Extended Data Tab.}

\begin{figure*}[ht]
\centering
\includegraphics[width=10cm]{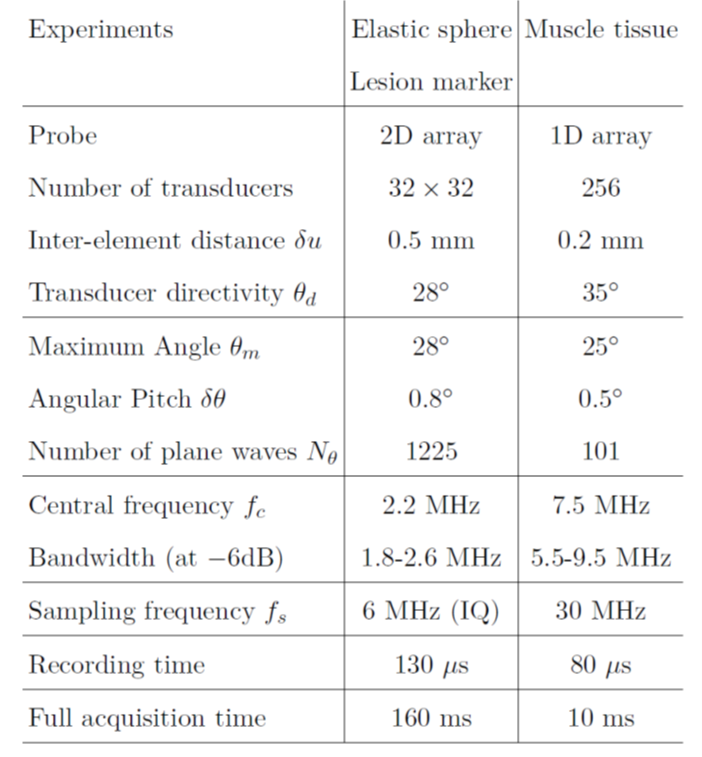}
\caption{\label{ProbeInfo} \textbf{Experimental parameters.}}
\end{figure*} 

\setcounter{figure}{0}
\renewcommand{\figurename}{Extended Data Fig.}

\begin{figure*}[ht]
\centering
\includegraphics[width=\textwidth]{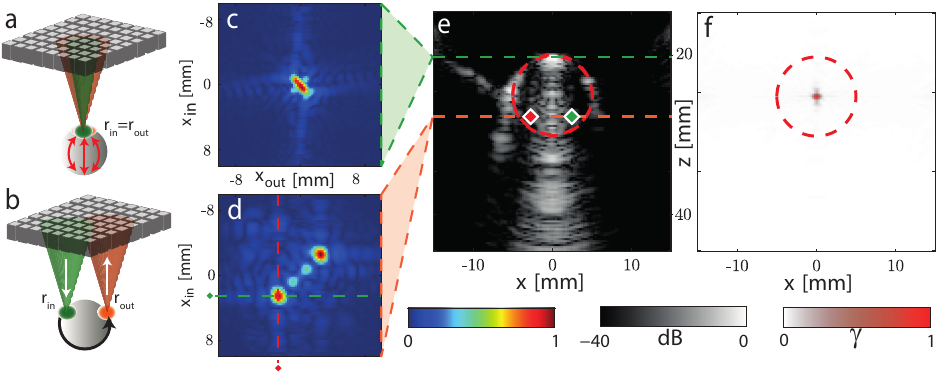}
\caption{\label{free} \textbf{{Elastic} target signature{s encoded in} the {reference} reflection matrix}. \textbf{a}, {The reference reflection matrix $\mathbf{R}_0$ is measured on the target sphere placed in water.} The confocal beamforming process applied to $\mathbf{R}_0$ selects {not only} the ballistic echo of the {sphere} but also its reverberations resulting from multiple reflections of bulk elastic waves (depicted by red arrows) at its inner surface. \textbf{b}, Matrix imaging decouples the input and output focal spots~\cite{Lambert2020a}, $\mathbf{r}_{\textrm{in}}$ and $\mathbf{r}_{\textrm{out}}$, to highlight the contribution of circumferential waves (depicted by a black arrow) generated by the incoming wave at a specific angle of incidence with respect to {sphere} surface~\cite{Prada1998}. \textbf{c}, Cross-section of the focused reflection matrix, {$\mathbf{R}_{0,xx}(y,z)$}, in the plane $y=0$ and at depth $z=21$ mm showing the diagonal contribution of the ballistic echo. \textbf{d}, Same matrix but at depth $z=28.5$ mm showing the off-diagonal contribution of circumferential waves. \textbf{e}, $(x,z)$-section of the confocal image in the plane $y=0$ showing the spatio-temporal dispersion of the target echo. \textbf{f}, Likelihood index map $\alex{\gamma}(\mathbf{r})$ (Eq.~\ref{eq P}) built from the fingerprint operator {indicating that we can accurately locate the target inside the reference environment (the sphere surface is highlighted by a red dashed line in panels \textbf{e} and \textbf{f}). Since in this case $\mathbf{R}\equiv\mathbf{R}_0\alex{(\mathbf{r}_0)}$, this result serves as a consistency check for the formalism.}}
\end{figure*} 

\newpage

\noindent { \textbf{Methods}} 

\noindent {\textbf{Ultrasound scanners and probes.}}

In the {elastic} {sphere} and {lesion} marker experiments, the acquisition of the reflection matrix is performed using a 2D matrix array of transducers (Vermon) whose characteristics are provided in \short{Extended Data Tab.~\ref{ProbeInfo}}. The electronic hardware used to drive the probe was developed by Supersonic Imagine in the context of a collaboration agreement with the Langevin Institute.

{In the calf experiment, the in vivo ultrasound data set has been collected by the SuperSonic Imagine company on an healthy volunteer from which informed consent had been obtained. Before being put at our disposal, this data set has been previously fully anonymized following standard practice defined by Commission nationale de l'information et des libert\'{e}s (CNIL).} The acquisition was performed using a medical ultrafast ultrasound scanner (Aixplorer Mach-30, Supersonic Imagine, Aix-en-Provence, France). 
The skin is placed in direct contact of a linear array of transducers (SL15-4, Supersonic Imagine) whose characteristics are also provided in \short{Extended Data Tab.~\ref{ProbeInfo}}.\\


\noindent {\textbf{Acquisition of the reflection matrix.}}

In each case, the reflection matrix is acquired using a set of plane waves~\cite{Montaldo2009}. For each plane wave of angles of incidence $\bm{\theta}_{\textrm{in}}=(\theta_x,\theta_y)$, the time-dependent reflected wave field  $R(\mathbf{u}_{\textrm{out}},\bm{\theta}_{\textrm{in}},t)$ is recorded by each transducer $\mathbf{u}_{\textrm{out}}$. This set of wave-fields forms a reflection matrix acquired in the plane wave basis, $\mathbf{R}_{\mathbf{u}\bm{\theta}} (t)=\left [ R(\mathbf{u}_{\textrm{out}},\bm{\theta}_{\textrm{in}},t) \right ]$. Since the transducer and plane wave bases can be related by a simple Fourier transform at the central frequency, the array pitch $\delta u$ and probe size $\Delta u$ dictate the angular pitch $\delta \theta$ and maximum angle $\theta_{max}$ necessary to acquire a full reflection matrix in the plane wave basis (Extended Data Tab.~\ref{ProbeInfo}). A set of plane waves are thus generated by applying appropriate time delays {$\Delta \tau (\bm{\theta}_\textrm{in},\mathbf{u}_{\textrm{in}})$} to each transducer $\mathbf{u}_{\textrm{in}}=(u_x,u_y)$ of the probe:
\begin{equation}
{\Delta \tau (\bm{\theta}_\textrm{in},\mathbf{u}_{\textrm{in}})} = [ u_x \sin \theta_x+u_y \sin \theta_y ]  /{c_0}. 
\end{equation}
where $c_0$ is the assumed speed-of-sound. The angular pitch and range as well as the number of illuminations are reported for each experiment in \short{Extended Data Tab.~\ref{ProbeInfo}}.
\vspace{5mm}

\noindent {\textbf{Focused beamforming of the reflection matrix.}} 

To form an image from the measured reflection matrix, a beamforming procedure has to be applied in emission and reception in order to generate a synthetic focusing on each focal point. In the time domain, a delay-and-sum beamforming process can be performed by applying appropriate time delays to the recorded signals~\cite{Bureau2023}. In the frequency domain, matrix products can be applied to project $\mathbf{R}$ in the focused basis~\cite{Lambert2020a}. The projection of the $\mathbf{R}-$matrix in the focused basis can be performed at each depth $z$ by means of the following matrix product:
\begin{equation}
{\mathbf{R}}_{\bm{\rho\rho}} (z,f) 
=  \mathbf{G}_{\mathbf{u}\bm{\rho}}^\dag (z,f) \times {\mathbf{R}}_{\mathbf{u}\boldsymbol{\theta}}(f)\times \mathbf{P}_{\boldsymbol{\theta} \bm{\rho} } ^{*}(z,f).
\label{projRrr}
\end{equation}
where $\mathbf{P}_{\boldsymbol{\theta} \bm{\rho}}(z,f)=\left [P\left(\boldsymbol{\theta},\bm{\rho},z,f \right) \right]$ and $\mathbf{G}_{\mathbf{u} \bm{\rho}}(z,f)=\left [ G \left(\mathbf{u},\bm{\rho},z,f \right) \right ]$ are the free space propagation matrices from the focused basis ($\bm{\rho}$) at depth $z$ to the plane wave 
($\bm{\theta}$) and transducer ($\mathbf{u}$) bases, respectively. Assuming an homogeneous speed of sound $c_0$, the coefficients of $\mathbf{P}_{\boldsymbol{\theta}\bm{\rho}}$ are expressed as follows
\begin{equation}
P (\boldsymbol{\theta}, \bm{\rho},f) = \exp\left[ i k \left (x \sin \theta_x + y \sin \theta_y +z \sqrt{1-\sin^2 \theta_x - \sin^2 \theta_y} \right)\right ].
\end{equation}
with $k=2 \pi f /c_0 $, the wave number. The coefficients of $\mathbf{G}_{\mathbf{u} \bm{\rho}}(z,f)$ are the 2D or 3D Green's function of the wave equation in homogeneous media:
\begin{subequations}
\begin{align}
{G}_{2D} (\mathbf{u},\bm{\rho},z,f) &= \frac{-i}{4} \mathcal{H}^{(1)}_0\left( k \sqrt{||\bm{\rho}-\mathbf{u}||^2+z^2} \right),
\\
{G}_{3D} ( \mathbf{u},\bm{\rho},z,f) &= -\frac{\exp\left(i k \sqrt{||\bm{\rho}-\mathbf{u}||^2+z^2} \right)}{4\pi \sqrt{||\bm{\rho}-\mathbf{u}||^2+z^2}}.
\end{align}
\end{subequations}
where $\mathcal{H}^{(1)}_0$ is the Hankel function of the first kind. Each coefficient of ${\mathbf{R}}_{\bm{\rho\rho}}(f)=[{R}(\bm{\rho}_{\textrm{out}},\bm{\rho}_{\textrm{in}},f)]$ is the response between a virtual source at point $(\bm{\rho}_{\textrm{in}},z)$ and a virtual detector at $(\bm{\rho}_{\textrm{out}},z)$ at frequency $f$.\\

In order to retrieve the axial resolution provided by the broadband feature of ultrasonic signals, a broadband focused reflection matrix $\overline{\mathbf{R}}_{\bm{\rho \rho}}(z)$ can be derived at each depth by coherently summing the monochromatic matrices over the frequency bandwidth:
\begin{equation}
\label{Rrr_focused_eq}
\overline{\mathbf{R}}_{\bm{\rho \rho}} (z) =\frac{1}{\Delta f} \int^{f_{+}}_{f_{-}} df {\mathbf{R}}_{\bm{\rho \rho}}  (z,f) ,
\end{equation}
where\ $f_{\pm}=f_c \pm \Delta f/2$ and $f_c$ is the central frequency. Each element of ${\mathbf{R}}_{\bm{\rho \rho}}(z)$ contains the signal that would be recorded by a virtual transducer located at\ $(\rout,z)$ just after a virtual source at\ $(\rin,z)$ emits a pulse of length $\delta t= \Delta f^{-1}$ at the central frequency $f_c$. 
A time-gated confocal image $\mathcal{I}(\bm{\rho},z)$ can be extracted from the diagonal of the broadband focused $\mathbf{R}-$matrix, such that:
\begin{equation}
\label{confocal}
\mathcal{I}(\bm{\rho},z)=\left |\overline{R}(\bm{\rho},\bm{\rho},z) \right|^2.
\end{equation}
The resulting confocal images are displayed in Figs.~\ref{config}d, Fig.~\ref{result}, Fig.~\ref{lesion}c and Fig.~\ref{muscle}b for the different experiments described in the paper.
\vspace{5mm}

\short{\noindent{\textbf{Highlighting the {elastic} target signature}}} \\

\short{{Even if the} response of {an elastic target} may be {spectrally and spatially quite complex~\cite{Thomas1994,Prada1998,Aubry2006,Robert2009,Gespa1987}, all these features can be adequately captured by measuring the target's reflection matrix. We illustrate this in} Extended Data Fig.~\ref{free} by showing the analysis of the {reference} reflection matrix associated with the 10-mm-diameter {metal} {sphere} immersed in water {without the scattering glass bead packing} (Supplementary Figure S4). In the following, we will refer to this reference (fingerprint) matrix as $\mathbf{R}_{0}(t,\mathbf{q}_0)$. The vector $\mathbf{q}_0=\lbrace \mathbf{r}_0,d \rbrace $ here accounts for the target state in this experiment, \textit{i.e.} its position $\mathbf{r}_0=(0,0,25)$ mm and diameter $d=10$ mm. Extended Data Figure~\ref{free}e displays the confocal image of the {sphere} obtained via the confocal beamforming algorithm described above and applied here to $\mathbf{R}_{0}${, which} highlights the complex signature of the target echo. Besides a specular component induced by the top of the {sphere}, this confocal image also shows a long temporal tail resulting from multiple reflections of bulk waves inside the {sphere} (Extended Data Fig.~\ref{free}a, Supplementary Section~S3). Beyond the confocal signal, the presence of circumferential {Rayleigh-like} waves propagating along the {sphere} surface~\citep{Royer1988,Clorennec2004} can be highlighted by decoupling the input and output focal spots and synthesizing the focused reflection matrix as described above (Extended Data Fig.~\ref{free}b, Methods). Cross-sections $\overline{\mathbf{R}}_{xx}(y,z)$ of the focused reflection matrix are displayed for the free space \rev{sphere} experiment in the mid-plane $y=0$ and at depths $z=21$ mm and 28.5 mm in Extended Data Figs.~\ref{free}c and d, respectively. While, at the first depth, the focused {$\mathbf{R}_0-$}matrix displays a predominant confocal signal characteristic of the specular echo on the {sphere} cap (Extended Data Fig.~\ref{free}c), off-diagonal echoes are observed at {greater} depth (Extended Data Fig.~\ref{free}d). As shown by previous works for a cylinder~\cite{Thomas1994,Prada1998} and confirmed by numerical simulations ({Supplementary Section~S3}), these echoes correspond to the following paths: (\textit{i}) excitation of a surface wave at an incident point $(\bm{\rho}_\textrm{in},z_0)$; (\textit{ii}) propagation of this circumferential wave along the {sphere} surface until the symmetric point $(\bm{\rho}_\textrm{out},z_0)=(-\bm{\rho}_\textrm{in},z_0)$ where it is back-converted {to} a compressional wave in the surrounding fluid.}\\ 

\noindent \textbf{Fingerprint operator based on a calibration measurement}\\

To build the fingerprint operator, the first strategy is to start from the measurement of a free-space reflection matrix $\mathbf{R}_0$ associated with the target at a given position $\mathbf{r}_0$. In principle, this matrix can be used as the value of the fingerprint operator at position $\mathbf{r}_0$: $\mathbf{F}(\mathbf{r}_0) \equiv \mathbf{R}_0$. In that case, the operation described in {Eq. ~\ref{eq P}} can be seen as an adaptive filter. This kind of filter may be not the most adequate approach for detection and localization purposes. 

First, depending on the experimental configuration, this reference matrix might be not necessarily specific enough with respect to the target environment ({Supplementary Figure~S5}). In the {sphere} {localization} experiment, the ballistic echo of {each} {sphere} shall be removed in order for the fingerprint matrix not to be dominated by the strong specular echo generated by the interface of the granular medium in the original experiment (Fig.~\ref{config}d). To that aim, the reference matrix $\mathbf{R}_0$ is filtered in the time domain by means of a Heavyside filter (Supplementary Figure~S6), such that:
\begin{equation}
\label{specular_filter}
{R}'_0(\mathbf{u}_{\textrm{out}},\bm{\theta}_{\textrm{in}},t)={R}_0(\mathbf{u}_{\textrm{out}},\bm{\theta}_{\textrm{in}},t) H(t-\tau_{\textrm{in}}(\bm{\theta}_{\textrm{in}},\mathbf{r}_0)-\tau_{\textrm{out}}(\mathbf{u}_{\textrm{out}},\mathbf{r}_0)-\delta t)
\end{equation}
with $\delta t\sim 1/\Delta f$, the temporal resolution of the recorded signals. $\tau_{\textrm{in}}$ and $\tau_{\textrm{out}}$ are the expected times-of-flight for the ballistic wave from the probe to the cylinder surface. At output (transducer basis), the time-of-flight $\tau_{\textrm{out}}$ can be expressed as follows:
\begin{equation} 
   \tau_{\textrm{out}}(\mathbf{u},\mathbf{r}_0)=\frac{\sqrt{(x_0-u_x)^2+(y_0-u_y)^2+(z_0-d/2)^2}}{c_0}.
\end{equation}
At input (plane wave basis), {$\tau_{\textrm{in}}$} is given by 
\begin{equation}
    {\tau_{\textrm{in}}(\bm{\theta},\mathbf{r}_0)}=\left [ x_0 \sin \theta_x+ y_0 \sin \theta_y +(z_0-d/2) \sqrt{1 - \sin^2 \theta_x -\sin^2 \theta_y } \right]  /{c_0}. 
\end{equation}

Second, the transfer function of an adaptive filter is not flat. Both the temporal and spatial frequency spectrum of the target can thus be altered if we consider $\mathbf{F}(\mathbf{r}_0) \equiv \mathbf{R}_0$ in Eq.~\ref{eq P}. This would reduce both the axial and transverse resolution of the imaging method in the real space and decrease the contrast of the target in the $\gamma$-map of the target likelihood index. To optimize the target signal, an inverse filter could be applied by considering the inverse matrix of $\mathbf{R}'_0$ as the building block of the fingerprint operator $F(\mathbf{r}_0)$. However, this would be at the price of an extreme sensitivity with respect to experimental noise and a loss in terms of signal-to-noise ratio.

A good compromise between an adaptive and inverse filtering process can be reached by performing a spectral whitening of $\mathbf{R}'_0$. In practice, this can be done by performing the singular value decomposition of ${\mathbf{R}}'_0(f)$ at each frequency $f$ (Supplementary Figure~S8):
\begin{equation}
\label{svd}
{\mathbf{R}}'_0(f)=\sum _p \sigma_p(f) \mathbf{U}_p(f) \mathbf{V}_p^{\dag}(f)
\end{equation}
where $\mathbf{U}_p$ and $\mathbf{V}_p$ are the output and input singular vectors of $\mathbf{R}'_0$. While, for a point-like target, the reflection matrix would only exhibit a predominant singular value, the spectrum of ${\mathbf{R}}'_0(f)$ displays a continuum of singular values due to the target size and the different resonant modes supported by each {sphere} {(Supplementary Section~S3)}.

To ensure the robustness of Eq.~\ref{eq P} with respect to experimental noise, only the eigenstates associated with the largest singular values shall be kept. In {practice}, the rank $N_S$ of the fingerprint operator is arbitrarily chosen such that the singular value ratio $\sigma_i/\sigma_1$ is larger than 0.4 {(Supplementary Figure~S8)}. The fingerprint operator is then obtained by whitening the singular value spectrum of this signal subspace such that:
\begin{equation}
\label{finger}
\mathbf{F}(\mathbf{r}_0,f)=\sum_{p=1}^{N_S} \mathbf{U}_p(f) \mathbf{V}_p^{\dag}(f) 
\end{equation}

\subsection*{Virtual shift of the target}

The next step is to deduce the spatial evolution of the fingerprint operator from its value at $\mathbf{r}_0$. This is done by performing a virtual translation of the target on each point $\mathbf{r}$ of the field-of-view. To that aim, the fingerprint matrix shall be projected at output in the plane wave basis:
\begin{equation}
    {\mathbf{F}}_{\bm{\theta}\bm{\theta}}(\mathbf{r}_0,f) = \mathbf{P}_{\mathbf{\theta}\mathbf{u}} (f)\times {\mathbf{F}}_{\mathbf{u}\mathbf{\theta}}(\mathbf{r}_0,f)
\end{equation}
From the plane wave basis, the target can be virtually shifted to any position $\mathbf{r}=\mathbf{r}_0+\Delta \mathbf{r}$ by the application of a shift operator $\mathbf{S}(\Delta \mathbf{r})$ at input and output of the matrix $\mathbf{F}_{\bm{\theta}\bm{\theta}}(\mathbf{r}_0,f)$. Mathematically, it can be written as the following matrix product:
\begin{equation}
    {\mathbf{F}}_{\bm{\theta}\bm{\theta}}(\mathbf{r},f) = \mathbf{S}(\Delta \mathbf{r},f) \circ {\mathbf{F}}_{\bm{\theta}\bm{\theta}}(\mathbf{r}_0, f)  \circ\mathbf{S}(\Delta \mathbf{r},f) 
\end{equation}
where the symbol $\circ$ stands for the Hadamard (element wise) product. In terms of matrix coefficients, the previous equation writes
\begin{equation}
\label{BF}
    {{F}}(\bm{\theta}_{\textrm{out}},\bm{\theta}_{\textrm{in}},\mathbf{r},f) = {S}(\bm{\theta}_{\textrm{out}},\Delta \mathbf{r},f) {{F}} (\bm{\theta}_{\textrm{out}},\bm{\theta}_{\textrm{in}},\mathbf{r}_0,f) {S}(\bm{\theta}_{\textrm{in}},\Delta \mathbf{r},f)
\end{equation}
where the coefficients $S(\bm{\theta},\Delta \mathbf{r},f)$ of the shift operator $\mathbf{S}(\Delta \mathbf{r},f)$ write
\begin{equation}
    S(\bm{\theta},\Delta \mathbf{r},f) = A(\bm{\theta},\mathbf{r})P(\bm{\theta},\mathbf{r},f)P^*(\bm{\theta},\mathbf{r}_0,f)=A(\bm{\theta},\mathbf{r})P(\bm{\theta},\Delta \mathbf{r},f)
\end{equation}
where $A(\bm{\theta},\mathbf{r})$ is an apodization factor that limits the angular range of the synthetic aperture at emission and reception which depends on whether the plane wave can reach the target at point $\mathbf{r}$ or not {(Supplementary Figure~S9)}. 

At last, the fingerprint operator can be projected back onto the acquisition basis of the recorded reflection matrix $\mathbf{R}$, such that
\begin{equation}
    \alex{\mathbf{F}}_{\mathbf{u}\bm{\theta}}(\mathbf{r},f) = \mathbf{P}_{\mathbf{\theta}\mathbf{u}}^{\dag}(f)\times \alex{\mathbf{F}}_{\bm{\theta}\bm{\theta}}(\mathbf{r},f)
\end{equation}

\subsection*{Numerical computation of the fingerprint operator}\label{num fingerprint op}

In the calf experiment, the fingerprint operator corresponds to a set of reflection matrices associated with a plane mirror of constant reflectivity characterized by a state $\mathbf{q}$ accounting for its position $\mathbf{r}$, orientation $\alpha$ and characteristic size $L$: $\mathbf{q}= \left \lbrace \mathbf{r}, \alpha, L \right \rbrace$ (Fig.~\ref{muscle}a). Each reference matrix $\mathbf{R}_{\mathbf{u}\bm{\theta}}(\mathbf{q},f)$ is computed numerically using the following matrix product:
\begin{equation}
  {\mathbf{R}}_0(\mathbf{q})=
    \mathbf{G}_{\mathbf{u} \mathbf{r}}(f) \times\boldsymbol{\Gamma}_{\mathbf{r}\mathbf{r}}(\mathbf{q})\times \mathbf{G}_{\mathbf{u} \mathbf{r}}^\top (f) \times   \mathbf{P}_{\boldsymbol{\theta} \mathbf{u} }^\dag (f) ,
\end{equation}
where $\boldsymbol{\Gamma}_{\mathbf{r}\mathbf{r}}(\mathbf{q})$ is a diagonal matrix whose diagonal coefficients $\gamma(\mathbf{r})$ stand for the reflectivity of the mirror in state $\mathbf{q}$. Note that, for such an object, the rank of the reflection matrix is equal to the number of resolution cells within the object, and that the $N_S$ non-zero singular values are degenerate~\cite{Robert2009}. The adaptive filter [$\alex{\mathbf{F}}(\mathbf{q})= \overline{  \alex{R}}_0(\mathbf{q})$]  or inverse filter operation [$\alex{\mathbf{F}}(\mathbf{q})= \alex{  \mathbf{R}}_0^{-1}(\mathbf{q})$] are therefore equivalent in this specific case.
\vspace{5mm}

\subsection*{Computational insights}

{In terms of computational efficiency, the best strategy is to calculate the likelihood index $\gamma$ by considering the reflection matrix expressed in the plane wave basis. The computation of $\gamma$ is then equivalent to a confocal beamforming process applied to the Hadamard (element-wise) product between the reflection matrix $\mathbf{R}_{\theta \theta}(f)$ and the fingerprint operator $\mathbf{F}_{\theta \theta}(f)$ (Supplementary Section~S7). Compared to a standard confocal beamforming process, the computational burden of our approach only consists in performing this Hadamard product between two matrices. For the 3D imaging experiments, such a product is performed in 3.5 s with Matlab (R2021a). The duration of this operation is almost negligible compared to the 3D confocal beamforming process that takes 1.3 min on a working station with 2 processors @2.20GHz, 128Go of RAM, and a GPU with 48 Go of dedicated memory~\cite{Bureau2023}. The computation time and power required to map the likelihood index $\gamma$ in Figs.~\ref{result} and \ref{lesion} is therefore almost equivalent to the beamforming of a confocal image.}  

{For the 2D muscle experiment (Fig.~\ref{muscle}), the Hadamard product between $\mathbf{R}_{\theta \theta}(f)$ and $\mathbf{F}_{\theta \theta}(f)$ is performed in 30 ms while the 2D confocal imaging process takes 80 ms on the same workstation. However, this operation shall be repeated for each couple $(\alpha,L)$. The number of tested angles $\alpha$ and correlation lengths $L$ being of 50 and 20, respectively, in the muscle experiment, the overall process thus takes approximately 10 s.}


\vspace{5mm}


\vspace{5mm}

\noindent\textbf{Data availability.} The ultrasound data generated in this study are available at Zenodo~\cite{LeBer2024} (\href{https://zenodo.org/records/14845780}{https://zenodo.org/records/14845780}).\\

\noindent \revv{\noindent\textbf{Code availability.} 
Codes used to post-process the optical data within this paper are
available from the corresponding author.}\\

\clearpage 

\noindent\textbf{Acknowledgments.}
The authors wish to thank the Somatex Company for providing the lesion marker. The authors are grateful for the funding provided by the European Research Council (ERC) under the European Union's Horizon 2020 research and innovation program (grant agreement no. 819261, REMINISCENCE project, A.A.). This project has also received funding from Labex WIFI (Laboratory of Excellence within the French Program Investments for the Future; ANR-10-LABX-24 and ANR-10-IDEX-0001-02 PSL*, M.F.) and from Agence Nationale de la Recherche (Grant No. 
ANR-22-ASTR-0020 under the AquaMat project, A.A.). L.M.R. was supported by the Austrian Science Fund (FWF) through Project no. \rev{P32300-N27} (WaveLand).\\

\noindent\textbf{Author Contributions Statement.}
A.A. and S.R. initiated the project. A.A. supervised the project. A.L.B. and X.J. designed and performed the experiments on the granular medium. A.L.B. performed the experiment on the lesion marker experiment. A.L.B. developed the post-processing tools for the target detection experiment. W.L. performed the muscle tissue experiment. A.G. developed the post-processing tools for the muscle tissue experiment. A.L.B., A.G., L.M.R., S.R. and A.A. developed the concept of the fingerprint operator and performed the theoretical study. A.L.B. and A.G. prepared the figures.  A.L.B., A.G. and A.A. prepared the manuscript. A.L.B., A.G., L.M.R., W.L., X.J., M.F., A.T., S.R. and A.A. discussed the results and contributed to finalizing the manuscript. \\

\noindent\textbf{Competing interests.}
A.L.B., A.G., L.M.R., W.L., X.J., M.F., A.T., S.R. and A.A. are inventors on a french patent related to this work held by Supersonic Imagine and CNRS (no. FR2314789, filed December 2023). M.F. is cofounder of the
SuperSonic Imagine company, which is commercializing one of the ultrasound platforms used in this study. W.L. is an employee of this company.


\clearpage

\renewcommand{\thetable}{S\arabic{table}}
\renewcommand{\thefigure}{S\arabic{figure}}
\renewcommand{\theequation}{S\arabic{equation}}
\renewcommand{\thesection}{S\arabic{section}}
\renewcommand{\thesubsection}{S\arabic{section}.\arabic{subsection}}
\renewcommand{\figurename}{Fig.}

\setcounter{equation}{0}
\setcounter{table}{0}

\setcounter{page}{1}

\newpage

\counterwithin*{figure}{part}
\counterwithin*{table}{part}

\stepcounter{part}

\section*{Table of Contents}

\begin{itemize}
\item Section~\ref{characterization}. Characterization of the {dense} granular {suspension}
\item Section~\ref{freebead}. {Free-space reflection matrix}
\item Section~\ref{complex}. {Back-scattered wave-field by an elastic {sphere}}
\item Section~\ref{contrast}. Theoretical prediction of the target contrast
\item Section~\ref{Cramer}. {Precision of the localization process}
\item Section~\ref{PFA}. {Detection threshold of potential targets}
\item Section~\ref{supp_link}. {Link between the mapping of the likelihood index and a confocal beamforming process}
\end{itemize}

\clearpage

\section{Characterization of the {dense} granular {suspension}}
\label{characterization}
\begin{figure*}[h!]
\centering
\includegraphics[width=12cm]{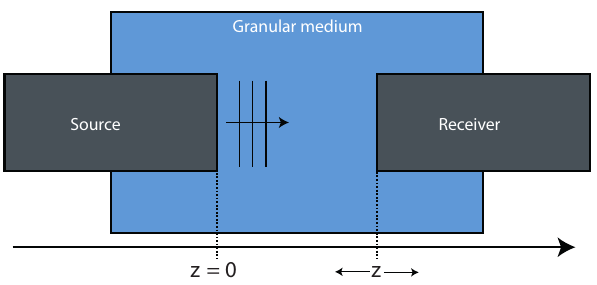}
\caption{\label{exp_config}\textbf{Measuring the scattering mean free path $\ell_s$ in the granular material.} Experimental set up dedicated to the measurement of the coherent wave across the granular medium.  An acoustic wave generated by a transducer disk propagates through the medium and is recorded by a similar transducer disk centered on the first one and placed at a distance $z$ from it.}
\end{figure*}

To measure the extinction length of the \corr{ballistic or direct wave}, a transmission configuration can be considered~\cite{Page1996}(Fig.~\ref{exp_config}). Two transducers (Olympus, model V382-SU ) are used to measure the coherent wave across the medium (circular aperture, diameter 12.7 mm, central frequency
3.5 MHz, bandwidth at -6 dB of the order of 70\%). One is working as a source and emits a quasi incident plane wave :
\begin{gather}
    \label{eq_ch_CaractTransmissionMilieuGranulaire_eCOS}
    \psi_{\textrm{inc}}(t) = \psi_0 \cos(2\pi f_c t) \exp \left [-{t^2}/(2\sigma^2) \right ] 
\end{gather}
with $f_c=3$ MHz and $\sigma=1/(2f_c)$.
The other transducer is used a receiver. The two transducers are mounted opposite each other along a common central axis.
The use of a wide transducer at reception is justified by the nature of the transmitted wave-field $\psi$ that is made of a coherent component $\left \langle \psi \right \rangle$ that resists to averaging over disorder and a multiple scattering coda $\delta \psi(\bm{\rho},z,t)$ wave that should vanish upon averaging  over a number of independent disorder configurations ($\bm{\rho}$ here accounts for the lateral position with respect to the $z-$axis). Taking advantage of spatial ergodicity, the coherent wave-field can be estimated by averaging the transmitted wave-field: (\textit{i}) over the transducer area; (\textit{i}) on different acquisitions separated by a mixing phase of the granular medium. The multiple scattering `coda' forms a speckle pattern
with each grain having an area comparable to the squared wavelength. Consequently, by choosing a circular transducer for reception whose active surface is 12.7 mm in diameter, the acquired signal is therefore averaged over around 500 speckle grains at 3 MHz which gives a reliable estimator of the coherent wave.

The receiver is mounted on a translation stage, which makes it possible to automate the measurement of the coherent wave for many distances $z$. In practice, we chose to vary $z$ from a few millimetres to around 40 mm, in steps of 0.1 mm, corresponding to almost 400 measurement points. To reduce electronic noise, the signals transmitted after many 2048 successive wave trains have been summed. A spatio-temporal window is then applied to the ultrasound data, so as to eliminate the multiple reflections between the transducers. A Fourier transform is finally performed in order to extract the Fourier-dependence of the coherent wave $\left \langle \psi \right \rangle (z,f) $. 
\begin{figure*}[h!]
\centering
\includegraphics[width=\textwidth]{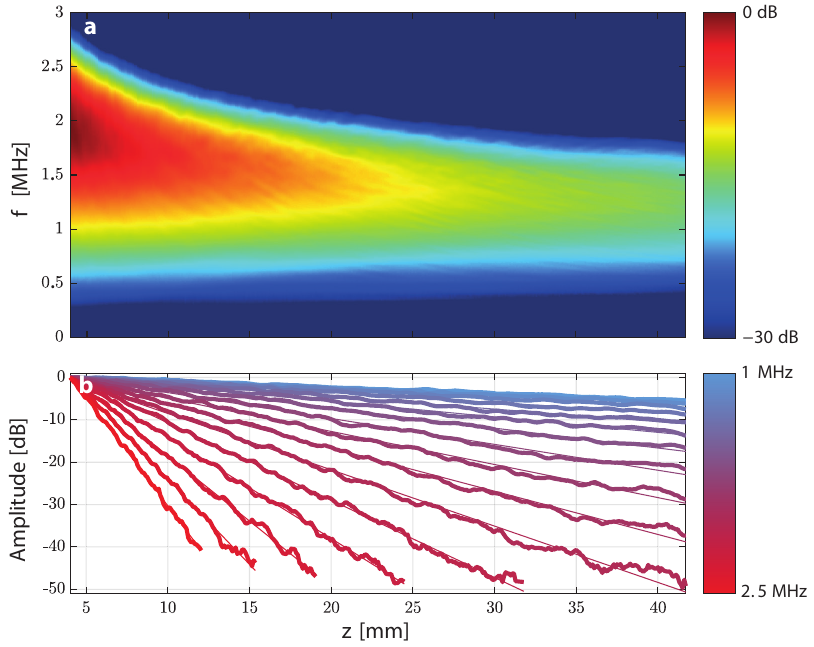}
\caption{\label{signal_decrease} \textbf{Depth evolution of the frequency spectrum of the coherent wave $\langle \psi( z,f) \rangle$.} \textbf{a}, Depth-frequency dependence of $\left |\langle \psi( z,f) \rangle \right |$. \textbf{b}, Depth dependence of  $\left |\langle \psi( z,f) \rangle \right |$ normalized by its value at $z=0$ at each frequency. The result is shown for a set of frequencies whose values are encoded with the color scale displayed on the right of the panel.}
\end{figure*}

Theoretically, the coherent wave-field can be expressed as follows:
\begin{align}
    \left \langle \psi \right \rangle(z,f) = \psi_0(f) \exp \left (\frac{i 2\pi f }{c_{\phi}(f)}z - \frac{z}{2\ell_\textrm{ext}(f)}\right )
\end{align}
with $c_{\phi}$, the phase velocity and 
\begin{equation}
\label{ls}
\ell_{\textrm{ext}}=\left[\ell_{s}^{-1} + \ell_{a}^{-1}\right]^{-1},
\end{equation}
the extinction length, taking into account both the scattering losses characterized by a scattering mean free path $\ell_s$ and absorption losses quantified by a characteristic absorption length $\ell_{a}$. A measurement of $\ell_{\textrm{ext}}$ is then performed at each frequency $f$ by investigating the depth decay of $ \left|\left \langle \psi \right \rangle(z,f)\right |$ (Fig.~\ref{signal_decrease}a). A linear fitting of $ \ln \left ( \left|\left \langle \psi \right \rangle(z,f) \right | \right )$ provides a measurement of $\ell_{\textrm{ext}} $ at each frequency (Fig.~\ref{signal_decrease}b). 

\begin{figure*}[h!]
\centering
\includegraphics[width=10cm]{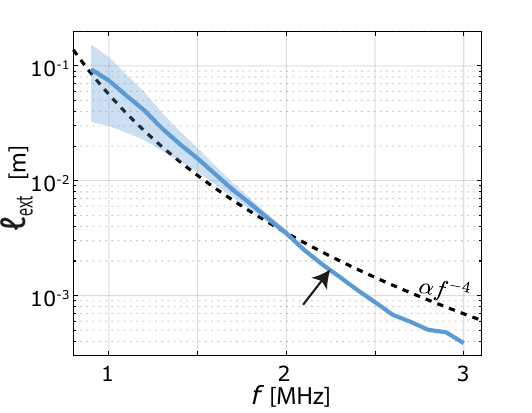}
\caption{\label{lext} \textbf{Measurement of the extinction length in granular media.} The measured extinction length $\ell_{\textrm{ext}}$ (blue line) is compared to the Rayleigh prediction (black dashed line). The black arrow indicates the scattering condition of the experiment depicted in Fig.~\ref{config}}
\end{figure*}

The result is displayed in Fig.~\ref{lext}. In the low frequency regime ($k_0a\ll 1$), the value of $\ell_{\textrm{ext}}$ follows a Rayleigh scattering law and is therefore extremely dispersive. At frequency $f=2.2$ MHz, the extinction length is $\ell_{\textrm{ext}}\simeq 1.5 $ mm (black arrow in Fig.~\ref{lext}).

The wave velocity $c_{\phi}$ has also been measured by investigating the phase of the coherent wave~\cite{LeBerThesis}. A value close to the speed-of-sound $c_0$ in water is found at the central frequency $f_c=2.2$ MHz: $c_{\phi}(f_c)\simeq c_0=1.55$ mm/$\mu$s.

At last, the temporal dependence of the ensemble average transmitted intensity~\cite{Page1995,Jia2004} has also been investigated to measure transport parameters in the granular medium in a higher frequency regime. Interestingly, such measurements provide a measurement of the absorption length $\ell_a\simeq 8$ mm~\cite{LeBerThesis} that we assume here as independent on frequency. Injecting the values measured for $\ell_a$ and $\ell_{\textrm{ext}}$ into Eq.~\ref{ls} has led to the following estimation for the scattering mean free path: $\ell_s \simeq 1.8$ mm at the central frequency $f_c=2.2$ MHz.
\clearpage

\section{Free-space reflection matrix}
\label{freebead}
 
\begin{figure*}[h!]
\centering
\includegraphics[width=10cm]{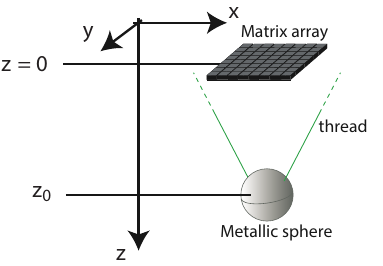}
\caption{\label{config2} \textbf{Experimental set up used to record the free space reflection matrix \corr{of the metal sphere}}. }
\end{figure*}

The experimental set-up used to record the free-space reflection matrix (Extended Data Fig.~\ref{free}) associated with each \rev{sphere} is displayed in Fig.~\ref{config2}. The \rev{sphere} is immersed in water and attached by two sewing threads glued on his surface. These threads can actually be seen in the ultrasound image shown in Extended Data Fig.~\ref{free}e.  

\begin{figure*}[h!t]
\centering
\includegraphics[width=14cm]{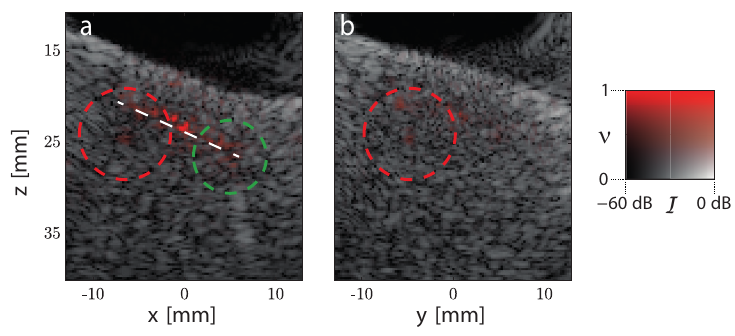}
\caption{\label{bad} \textbf{Detrimental effect of specular echoes on the likelihood index map.} Confocal image (B\&W scale) and superimposed $\gamma$-map (red scale) associated with the 10-mm diameter \rev{sphere}. The $\gamma$-map is built using the raw matrix $\mathbf{R}_0$ as the fingerprint operator. The frequency bandwidth is 1.5-2.5 MHz. \textbf{a}, $(x,z)-$ cross-section in the plane $y=y_1$. The white dashed line shows the artifact due the specular echoes on the granular medium surface. \textbf{b}, $(x,z)-$ cross-section in the plane $x=x_1$. The positions of \rev{sphere}s 1 and 2 are represented by red and green dashed lines, respectively.}
\end{figure*}

Figure~\ref{bad} shows the likelihood index map for \rev{sphere} 1 using the raw matrix ${R}_0$ as the fingerprint operator (${F}={R}_0$). This operator completely fails to reveal the presence of \rev{sphere} 1 because the direct echo of the \rev{sphere} is not specific with respect to its environment. The interface of the scattering medium actually generates a strong specular echo that accounts for the surintensity of the $\gamma-$map lying along the dashed white line in Fig.~\ref{bad}a. There is a vertical shift of 5 mm with respect to the real position of the interface in the confocal image that corresponds to the radius of \rev{sphere} 1. 

\begin{figure*}[h!t]
\centering
\includegraphics[width=14cm]{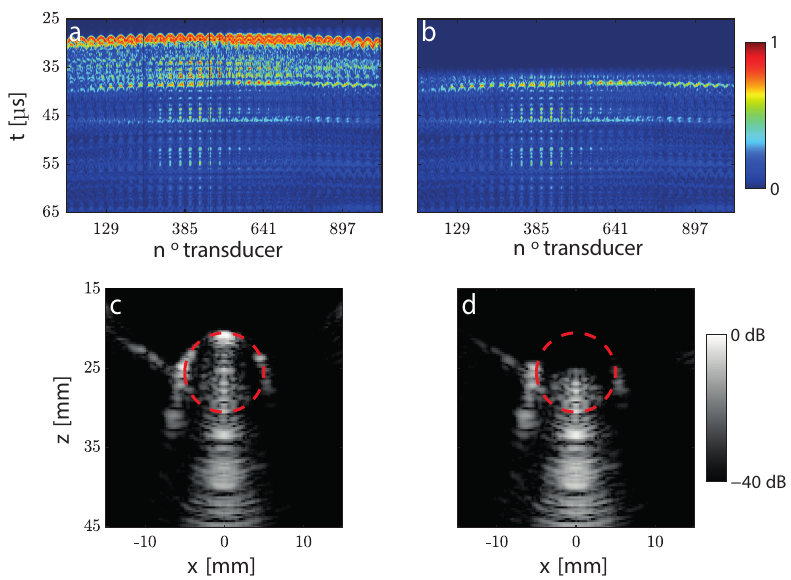}
\caption{\label{ballistic_filter}\textbf{Time filtering of the specular \rev{sphere} echo.} \textbf{a}, Time evolution of the reflected wave-field $R_0(\mathbf{u}_{\textrm{out}},\bm{\theta}_0,t)$ measured by the probe for the normal incident plane wave $\bm{\theta}_0=(0,0)$.
\textbf{b}, Time evolution of the corresponding filtered wave-field $R'_0(\mathbf{u}_{\textrm{out}},\bm{\theta}_0,t)$. \textbf{c}, $(x,z)-$section of the confocal image $\mathcal{I}$ (Eq.~\ref{confocal}) in the plane $y=y_1$ built from the free-space matrix $\mathbf{R}_0$. \textbf{c}, Same section of the confocal image $\mathcal{I}$ (Eq.~\ref{confocal}) built from the filtered matrix $\mathbf{R}'_0$. The data shown here are for \rev{sphere} 1.}
\end{figure*}

\begin{figure*}[h]
\centering
\includegraphics[width=\textwidth]{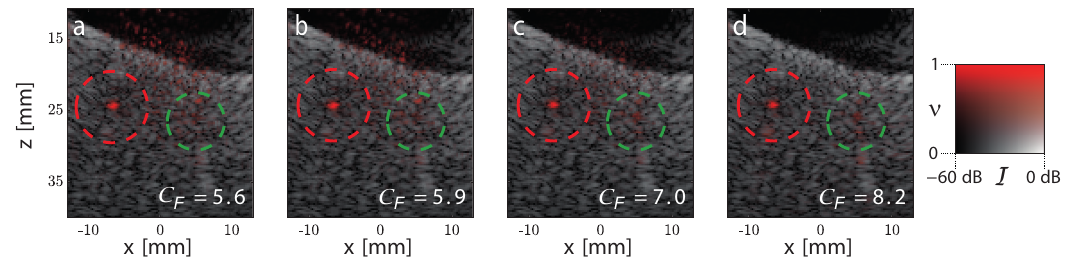}
\caption{\label{fingerprint_evolution} \textbf{Optimization of the fingerprint operator.} The $(x,z)-$cross-section ($y=y_1$) of the $\gamma-$map for \rev{sphere} 1 (red scale) superimposed to the confocal image (B\&W scale) is computed by considering as a fingerprint operator: \textbf{a}, the time filtered matrix ($\mathbf{F}=\mathbf{R}'_0$); \textbf{b}, the $N_1$ first eigenstates of $\mathbf{R}'_0$ $\left ( \mathbf{F}=\sum_{p=1}^{N_1} \sigma_p \mathbf{U}_p\mathbf{V}_p^{\dag} \right )$; \textbf{c},  after its singular spectrum whitening (Eq.~\ref{finger}); \textbf{d}, after the angular filtering described in Fig.~\ref{angular_filter}. The frequency bandwidth is here 1.5-2.5 MHz.}
\end{figure*}

As described in the Methods section, a time filtering of specular echoes should thus be performed on each free-space reflection matrix (Eq.~\ref{specular_filter}), in order to make the fingerprint operator more specific. Figure~\ref{ballistic_filter} illustrates the effect of this filter: (\textit{i}) on the raw signals by comparing the initial dataset before (Fig.~\ref{ballistic_filter}a) and after (Fig.~\ref{ballistic_filter}b) filtering of the specular \rev{sphere} echo; (\textit{ii}) on the confocal image extracted from the measured (Fig.~\ref{ballistic_filter}c) and filtered (Fig.~\ref{ballistic_filter}d) matrices, ${R}_0$ and ${R}'_0$, respectively. In the latter image, the specular echo that initially appeared on the cap of the \rev{sphere} in Fig.~\ref{ballistic_filter}c is completely removed in Fig.~\ref{ballistic_filter}d.  
The importance of this filter is highlighted by comparing the likelihood index $\gamma$ built from $\mathbf{R}'_0$ (Fig.~\ref{fingerprint_evolution}a) and its initial counterpart (Fig.~\ref{bad}a). \rev{sphere} 1 is now localized with a satisfying contrast $\mathcal{C}_\gamma=5.6$.   

\begin{figure*}[h]
\centering
\includegraphics[width=12cm]{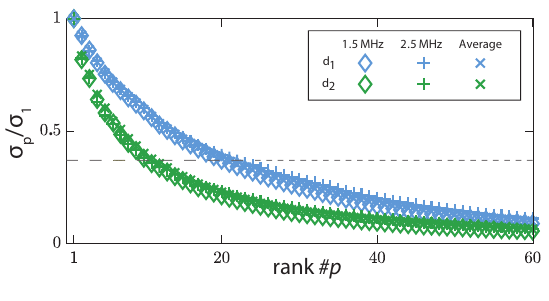}
\caption{\label{singular_values} \textbf{Singular values of the free-space and filtered matrices} $\mathbf{R}'_0(f)$ at frequency $f=1.5$ MHz and 2.5 MHz. Their values averaged over the frequency bandwidth is also displayed. The horizontal dashed line accounts for the threshold applied to build the fingerprint operator $\mathbf{F}$ in Eq.~\ref{finger}. Each singular value is normalized by its maximum value $\sigma_1$.}
\end{figure*}

To improve this contrast, a singular value decomposition can then be applied to the filtered matrix ${\mathbf{R}}'_0(f)$ at each frequency $f$ (Eq.~\ref{svd}). The resulting singular values are displayed in Fig.~\ref{singular_values}. Only the eigenstates associated with singular values checking $\sigma_i>0.4 \sigma_1$ can be considered to build the fingerprint operator. The resulting $\gamma-$map is displayed in Fig.~\ref{fingerprint_evolution}b. The contrast improvement is modest ($\mathcal{C}_\gamma=5.9$) but is clearly improved after whitening the singular value spectrum (Eq.~\ref{finger}).  The corresponding $\gamma-$map shown in Fig.~\ref{fingerprint_evolution}c displays a contrast $\mathcal{C}_\gamma=7.0$.

\begin{figure*}[ht]
\centering
\includegraphics[width=12cm]{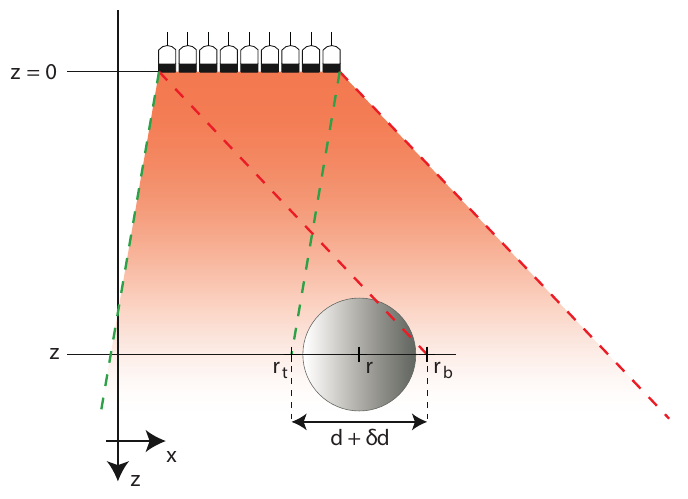}
\caption{\label{angular_filter} \textbf{Scheme describing the selection of plane waves kept in the fingerprint operator.} The orange area is a view of the selected angular sector. All the waves coming from outside this area are removed by the angular filter.}
\end{figure*}

A last improvement consists in an angular decomposition of the fingerprint operator in order to filter the contribution of plane waves that cannot come from the target. This operation is illustrated by Fig.~\ref{angular_filter} and its result is shown in Fig.~\ref{fingerprint_evolution}d. The final contrast $\mathcal{C}_\gamma$ reaches the excellent value of 8.2. Note that this value differs from the one reached in the accompanying paper (Fig.~\ref{result}a, $\mathcal{C}_\gamma$) because the frequency bandwidth differs: 1.5-3.5 MHz in Fig.~\ref{fingerprint_evolution}d  vs. 1.8-2.6 MHz in Fig.~\ref{result}a. In the former case, the frequency-averaged value of the scattering mean free path $\ell_s$ (Fig.~\ref{lext}) is larger than in the latter case.  
\clearpage

\section{Back-scattered wave-field by an elastic {sphere}}
\label{complex}
\begin{figure}[h!tb]\centering
    \includegraphics[scale=1]{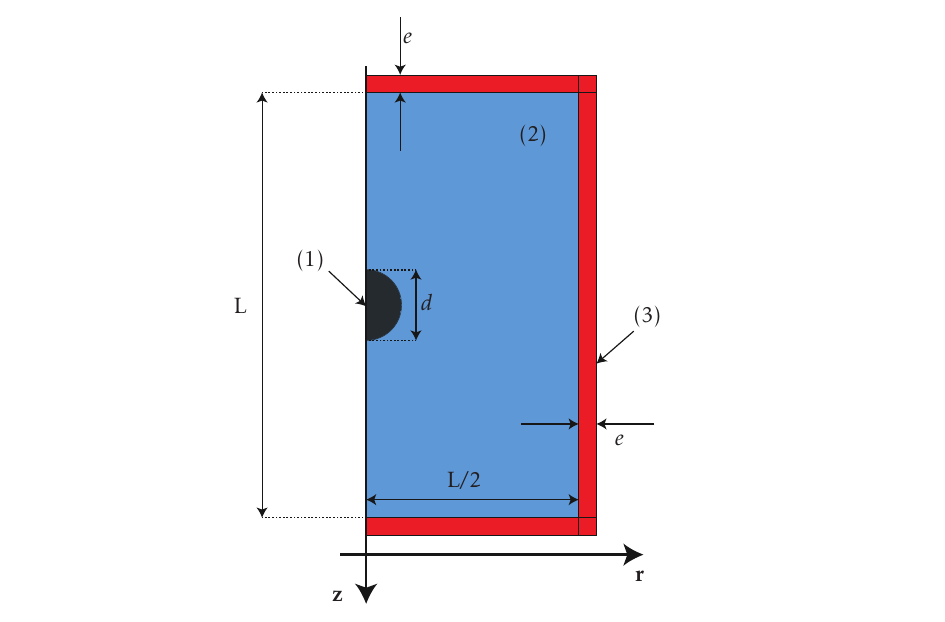}
    \caption{\textbf{Numerical simulation of the \rev{sphere} response.} Scheme depicting the different parts of the simulation domain.}
    \label{fig_ch_DIMO_SimuCOMSOLParams}
\end{figure}

A COMSOL Multiphysics 6.0 simulation has been used to compute the wave-field scattered by the \rev{metal} \rev{sphere}.  As the mechanical properties of the \rev{sphere} used in the experiment are not known exactly, we use tabulated values for the simulation (Tab.~\ref{tbl_ch_DIMO_ParamSIMUComsol}). We also took advantage of the rotation symmetry of the \rev{sphere} to use a two-dimensional axisymmetric simulation, which is less demanding in terms of resources and computing time. The aim of the simulation is first to estimate the field induced by the \rev{sphere} when illuminated by an incident plane wave. In a second step, we will simulate focusing by coherently summing different plane waves obtained from the rotation of the initial plane wave simulation. The geometry, dimensions and physical quantities used for the simulation are all compiled in Fig.~\ref{fig_ch_DIMO_SimuCOMSOLParams} and Tab.~\ref{tbl_ch_DIMO_ParamSIMUComsol}.

\begin{table}[h]\centering
    \begin{tabular}{c|c|c}
    & \textbf{Parameter} & \textbf{Value}  \\ \hline \hline
    \multirow{5}{*}{\textbf{Medium $(1)$}}     
    &Medium                    & Isotropic Solid \\ \cline{2-3}
    &$d$                               & \qty{10}{\milli\meter}  \\ \cline{2-3}
    &$\rho_b$                          & \qty{5700}{\kilo\gram\per\cubic\meter}  \\ \cline{2-3}
    &$c_\textrm{p}$                        & \qty{2700}{\meter\per\second}  \\ \cline{2-3}
    &$c_\textrm{t}$                        & \qty{2000}{\meter\per\second}  \\ \cline{1-3}
    \multirow{4}{*}{\textbf{Medium $(2)$}}     
    &Medium Type                    & Liquid  \\ \cline{2-3}
    &$L$                               & \qty{60}{\milli\meter}  \\ \cline{2-3}
    &$\rho_0$                          & \qty{1000}{\kilo\gram\per\cubic\meter}  \\ \cline{2-3}
    &$c_0$                             & \qty{1500}{\meter\per\second}  \\ \cline{1-3}
    \multirow{2}{*}{\textbf{Medium $(3)$}}     
    &Medium                   & PML  \\ \cline{2-3}
    &$e$                               & \qty{2,5}{\milli\meter}  \\ \cline{1-3}
    \multirow{4}{*}{\textbf{Simulation}}    
    &Incident wave        & Plane wave \\ \cline{2-3}
    &$f_\textrm{min}=\Delta f$             & \qty{6}{\kilo\hertz}  \\ \cline{2-3}
    &$f_\textrm{max}$                      & \qty{6}{\mega\hertz}  \\ \cline{2-3}
    &Frequency enveloppe           & Hann window
    \end{tabular}
    \caption{A set of parameters used to simulate the response of a \rev{metal sphere} immersed in water in the presence of an incident plane wave.}
    \label{tbl_ch_DIMO_ParamSIMUComsol}
\end{table}

\begin{figure}[h!tb]\centering
    \includegraphics[scale=1]{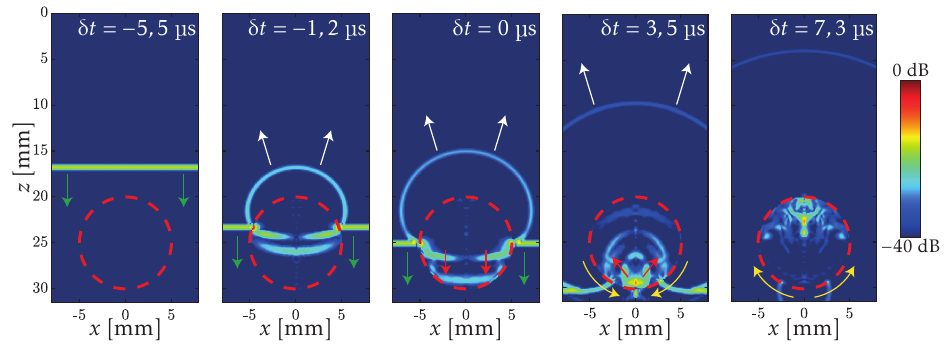}
    \caption{\textbf{Wave propagation movie for an incident plane wave}. Different snapshots are shown for an incident plane wave, observed in the $y=0$ plane. The incident wave propagates towards increasing $z$ (green arrows). Its interaction with the \rev{sphere} gives rise to a specular back-refection (white arrows). Bulk waves are also generated within the \rev{sphere} (red arrows). Finally, Rayleigh-type surface waves are also generated (yellow arrows). The time $\delta t=0$ corresponds to the passage of the plane wave through the $z=z_0$ plane (center of the \rev{sphere}).}
    \label{fig_ch_DIMO_SimuCOMSOLNoFocusOnePW}
\end{figure}

Comsol simulation gives us access to a wide range of physical quantities both within the \rev{sphere} and in the water. We have chosen to focus on the pressure in the water and the amplitude of the displacement field in the ball. These quantities are obtained for each frequency at any point in space. A Fourier transform is then computed after modulation by a Hann window to reconstruct a movie of wave propagation in the system in the plane $y=0$. Figure~\ref{fig_ch_DIMO_SimuCOMSOLNoFocusOnePW} shows the incident plane wave propagating through water until it is reflected by the \rev{sphere}. In addition to being directly reflected, part of the wave is transmitted into the \rev{sphere}, as two wave fronts emerge within it. The first, moving faster, corresponds to the longitudinal wave, while the second is associated with the transverse wave. In addition to these bulk waves, circumferential waves are also generated, corresponding to Rayleigh waves. These circulate along the surface and propagate without stopping, so that they can turn around the \rev{sphere}. 

From this mother simulation, it is also possible to simulate focusing at any point of the simulation grid using the coherent sum of numerous plane waves. To simulate a three-dimensional focus from the parent simulation in axi-symmetric two-dimensional space, we rely on the rotation invariance of the system. To that aim, an observation grid is first defined for the simulated quantities. The idea is then to apply to the coordinates of this grid the rotation operation required to give the orientation of the wave vector of the parent simulation to the wave vector we are trying to simulate. For each plane wave to be simulated, we thus define a new associated grid that can be expressed in the cylindrical coordinates of the parent simulation. On each of these grids, the quantities of interest are finally estimated by interpolation from the grid of the parent simulation. At this stage, we now know the frequency dependence of the quantities of interest for all incident plane waves in a common grid.

\subsection{Origin of the bright tail}

\begin{figure}[h!tb]\centering
    \includegraphics[scale=1]{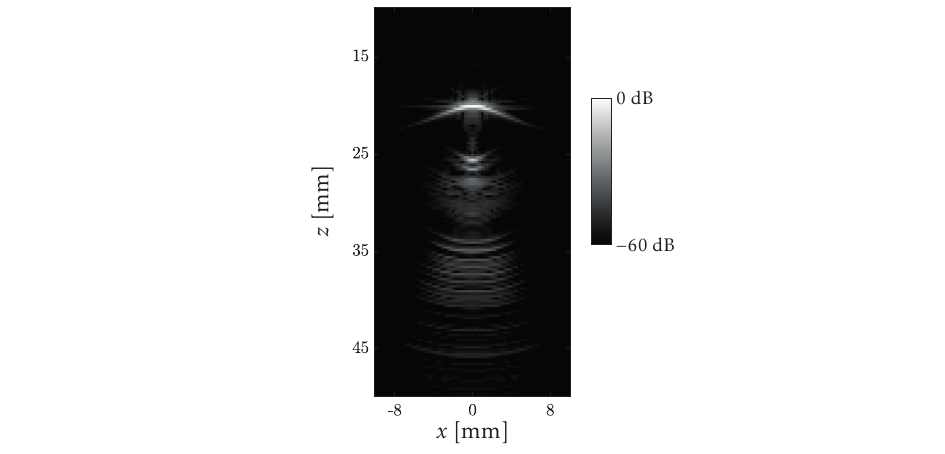}
    \caption{\textbf{Confocal image obtained from the simulated reflection matrix}. The $(x,z)-$section of the image is displayed in the plane $y=0$. }
    \label{fig_ch_DIMO_SimuCOMSOLImageConfocale}
\end{figure}

To form a confocal image from the parent simulation, the transducer coordinates of the matrix probe can be chosen as the acoustic pressure measurement grid. In this way, the reflection matrix of the simulated system can be computed. A confocal image can be deduced by applying a confocal beamforming process, as described in the Methods section and displayed in in Fig.~\ref{fig_ch_DIMO_SimuCOMSOLImageConfocale}. As already observed experimentally (Extended Data Fig.~\ref{free}e), the confocal image shows the \rev{sphere} cap but also a bright tail, whose origin can now be investigated.

A film of wave propagation can be formed when all plane waves are suitably delayed to sum coherently at a point belonging to the bright tail. We are interested, for example, in the pixel $\mathbf{r}_{\textrm{in}} = (0;0;{25.5})$ mm and we set up an observation grid in the $y=0$ plane, on which we are now able to know the frequency dependence of the quantities of interest for the various incident plane waves. In order to focus on $\mathbf{r}_{\textrm{in}}$, it remains to apply beamforming in the frequency domain (Methods) so that all the plane waves sum coherently in $\mathbf{r}_{\textrm{in}}$ at the ballistic time ($\delta t=0$). The wave propagation movie is finally obtained by summing the wave-field obtained for each plane wave at each point $\mathbf{r}$ and then calculating the inverse time Fourier transform after bandwidth modulation by a Hann window.

\begin{figure}[h!tb]\centering
    \includegraphics[scale=1]{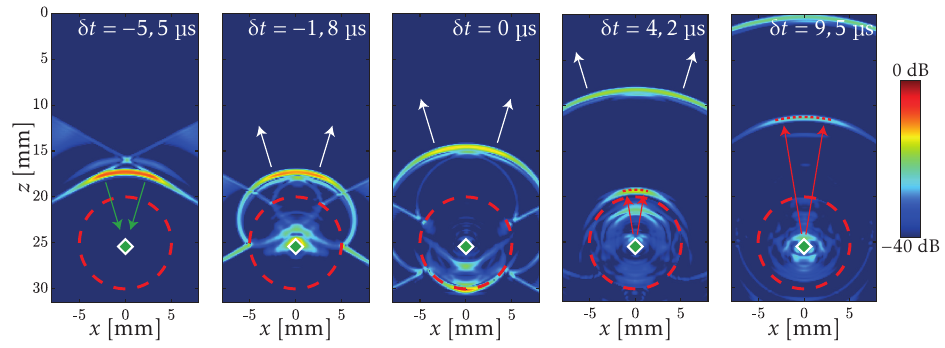}
    \caption{\textbf{Wave propagation movie for an incident focused wave}. Snapshots from the film of wave propagation in the system when focused on the point $\mathbf{r}_{\textrm{in}} = (0;0;{25.5})$ mm (green diamond) and observed in the plane $y=0$. The incident wave (green arrows) gradually converges towards $\mathbf{r}_{\textrm{in}} $. When the incident wave hits the \rev{sphere} surface, a fraction of the energy is reflected (specular reflection, white arrows). Another fraction of the energy is \corr{converted into a whispering mode} and then radiated back into the medium, (bulk reverberations, red arrows). The red dashed line represents the wave-front contributing to the reverberation observed in Fig.~\ref{fig_ch_DIMO_SimuCOMSOLImageConfocale} at point $\mathbf{r}_{\textrm{in}} $.}
    \label{fig_ch_DIMO_SimuCOMSOLFocusOnCenter}
\end{figure}

Snapshots of this movie are displayed in Fig.~\ref{fig_ch_DIMO_SimuCOMSOLFocusOnCenter}.  The incident focused wave does indeed appear to converge towards the point $\mathbf{r}_{\textrm{in}}$ as long as they propagate freely in the water. When they hit the surface of the ball, part of the energy is reflected, while another part is transmitted into the \rev{sphere}. Longitudinal and transverse waves seem to undergo a refocusing process beneath the \rev{sphere} surface and multiple reflections are observed at later time lapses. This process transfers a significant amount of energy back to the water, in the direction of the probe.

To form the confocal image, the echoes from $\mathbf{r}_{\textrm{in}}$  focus are precisely summed, selecting those corresponding to the supposed time delay between $\mathbf{r}_{\textrm{in}}$ and each transducer element. This echo highlighted by a red dashed line in Fig.~\ref{fig_ch_DIMO_SimuCOMSOLFocusOnCenter} corresponds here to the first reflection on the back of the \rev{sphere}. 
In the confocal image (Fig.~\ref{fig_ch_DIMO_SimuCOMSOLImageConfocale}), the bright tail thus exists thanks to the multiply-reflected bulk waves inside the \rev{sphere} generated by incicent waves intended to focus at these points and then radiated back towards the probe with the time delay expected in a homogeneous medium.

\subsection{Origin of off-diagonal echoes}

\begin{figure}[h!tb]\centering
    \includegraphics[scale=1]{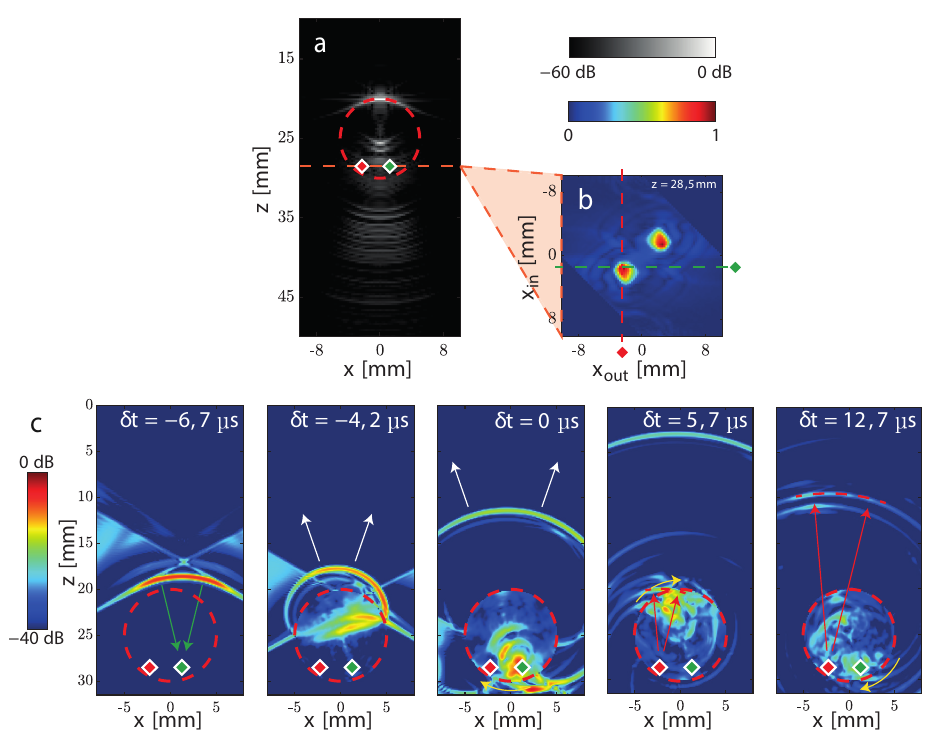}
    \caption{\textbf{Origin of off-diagonal echoes}. \textbf{a}, Confocal image in the $y=0$ plane. \textbf{b}, $x-$ cross-section of the focused reflection matrix in the focused basis in the plane $y_{\textrm{in}}=y_{\textrm{out}}=0$  and at depth $z=28.5$ mm. \textbf{c}, Snapshots of the wave propagation movie for an incident wave focusing on the point $\mathbf{r}_\textrm{in} = (1.25;0;28.5)$ mm (green diamond) and observed in the plane $y=0$. The incident wave-front (green arrows) gradually converges towards $\mathbf{r}_\textrm{in}$ . When the incident wave hits the surface, a fraction of the energy is reflected and gives rise to the specular echo (white arrows). Another fraction of the energy is \corr{converted into a circumferential surface} wave (yellow arrows) and is then radiated back towards the probe (red arrows). The red dashed lines represent the echoes in the $y=0$ plane actually used by the probe to focus on the output point $\mathbf{r}_\textrm{out}=(-\bm{\rho_{\textrm{in}}},z)$ (red diamond). The chosen couple of points $(\mathbf{r}_\textrm{in},\mathbf{r}_\textrm{out})$ corresponds to the strong off-diagonal echo displayed in \textbf{b}.}
    \label{fig_ch_DIMO_SimuCOMSOLFocusOnSide}
\end{figure}

To better understand the origin of off-diagonal signal in the focused reflection matrix (Extended Data Fig.~\ref{free}d), we can once again rely on the simulation described above. Focused beamforming can be applied at input and output of the simulated reflection matrix in order to synthesize the focused reflection matrix at different depths (Method). As observed experimentally (Extended Data Fig.~\ref{free}d), we retrieve for certain depth strong off-diagonal echoes (Fig.~\ref{fig_ch_DIMO_SimuCOMSOLFocusOnSide}).

Remarkably, the simulation once again allows us to track the origins of this off-diagonal echo, by simulating an incident wave focusing on the corresponding point $\mathbf{r}_{\textrm{in}}$. Snapshots of the corresponding movie are shown in Fig.~\ref{fig_ch_DIMO_SimuCOMSOLFocusOnSide}c. It shows once again the incident wave converging towards the point $\mathbf{r}_{\textrm{in}}$ before hitting the \rev{sphere} surface. A predominant surface wave is generated and propagates around the \rev{sphere} (yellow arrow). The wave-front contributing to the off-diagonal echo in Fig.~\ref{fig_ch_DIMO_SimuCOMSOLFocusOnSide}b is highlighted by a red dashed line. It indeed corresponds to the echo produced by the circumferential wave. 

\section{Theoretical prediction of the target contrast}
\label{contrast}

In the first experiment described in the accompanying paper (Fig.~\ref{config}), the recorded reflection matrix $\mathbf{R}$ can be decomposed into a target component $\sigma_T \mathbf{R}_0$, and a multiple scattering contribution $\mathbf{M}$:
\begin{equation}
\mathbf{R}=\sigma_T \mathbf{R}_0+\mathbf{M}
\end{equation}
These two matrices are considered as fully uncorrelated. The single scattering contribution of the granular medium is neglected since multiple scattering is strongly predominant at the targets' depths in Fig.~\ref{config}. We will consider a normalized matrix $\mathbf{R}_0$ such that $\sigma_T$ accounts for the target signal: $$\left \langle |R_0(\mathbf{u}_\textrm{out},\bm{\theta}_{\textrm{in}},f) |^2\right \rangle \equiv 1,$$ 
where the symbol $\langle \cdots \rangle$ stands for ensemble average.

\subsection{Target contrast in confocal imaging}

Injecting Eq.~\ref{projRrr} and \ref{Rrr_focused_eq} into Eq.~\ref{confocal} leads to the following expression for the mean confocal intensity:
\begin{equation}
\label{mean_intensity}
\mathcal{I} = \left \langle \left | \frac{1}{\Delta f}\int_{f_-}^{f_+} df \sum_{\bm{\theta}_{\textrm{in}}}^{N_{\theta}}\sum_{\mathbf{u}_\textrm{out}}^{N_u} R(\mathbf{u}_\textrm{out},\bm{\theta}_{\textrm{in}},f) P^*(\bm{\rho},\bm{\theta}_{\textrm{in}},f)G^*(\bm{\rho},\mathbf{u}_{\textrm{out}},f)\right |^2 \right \rangle.
\end{equation}

In the multiple scattering regime, confocal beamforming is an incoherent process:  Each term in the triple sum of Eq.~\ref{mean_intensity} can be seen as a random phasor. The corresponding multiple scattering contribution can be rewritten as follows:
\begin{equation}
\label{mean_intensityM}
\mathcal{I}_M = \frac{\delta f}{\Delta f}\sum_{\bm{\theta}_{\textrm{in}}}^{N_{\theta}}\sum_{\mathbf{u}_\textrm{out}}^{N_u} \left \langle \left | M(\mathbf{u}_\textrm{out},\bm{\theta}_{\textrm{in}},f) \right |^2 \right \rangle
\end{equation}
where $\delta f$ is the correlation frequency of the multiple scattering noise. Let $\sigma_M^2=\left \langle \left | M(\mathbf{u}_\textrm{out},\bm{\theta}_{\textrm{in}},f) \right |^2 \right \rangle$ be the power of multiple scattering noise recorded by the probe. The mean multiple scattering intensity thus scales as follows:
\begin{equation}
\mathcal{I}_M \sim \frac{N_u N_{\theta}}{N_f}  \sigma_M^2
\end{equation}
with $N_f=\Delta f /\delta f$, the number of independent frequency grains in the frequency bandwidth. The confocal beamforming process amounts to increase the multiple scattering intensity by a factor $N_{\theta}$ due to the beamforming at input, by a factor $N_u$  due to the beamforming at output. Moreover, the different frequency components of the wave-field do not sum coherently which leads to a decrease of the multiple scattering intensity by the number $N_f$ of independent frequency grains in the bandwidth.

To derive an equivalent scaling for the target intensity, we will first consider the case of a point-like target. In that case, confocal beamforming is a perfectly coherent process with respect to the singly-scattered echo of the target: Each term in the triple sum of Eq.~\ref{mean_intensity} adds constructively. The associated confocal intensity $\mathcal{I}_P $ thus scales as follows:
\begin{equation}
\mathcal{I}_P \sim {N_{\theta}^2 N_u^2}\sigma_T^2
\end{equation}
 The confocal beamforming process amounts to increase the point-like target intensity $\mathcal{I}_P$ by a factor $N_{\theta}^2$ due to the focused beamforming at input and by a factor $N_u^2$  due to the focused beamforming at output.

For a more complex target, the phasors in Eq.~\ref{mean_intensity} are only partially coherent. This case is therefore intermediate between the multiple scattering component and the point-like target case. At each frequency, only one eigenstate of the reflection matrix, its specular component, leads to a coherent sum in the confocal beamforming process. If we assume that the target matrix exhibits a step-like distribution of singular values with $N_S$ the number of non-zero singular values, the confocal intensity should thus be decreased by a factor $N_S$ compared to the case point-like target. Moreover, the target exhibits a partially incoherent frequency spectrum. The associated intensity $\mathcal{I}_T $ will be also decreased by a factor $N_T$ compared to the point-like target case, $N_T$, being the number of independent coherence grains in the frequency bandwidth. The scaling $\mathcal{I}_T $ is therefore as follows:
\begin{equation}
\mathcal{I}_T \sim \frac{\mathcal{I}_P }{N_S N_T} \sim \frac{N_{\theta}^2N_u^2}{N_S N_T} \sigma_T^2
\end{equation}
where $N$ is the number of transducers. $N_S$ and $N_T$ are the number of spatial and temporal degrees of freedom exhibited by the target. 

In the diffusive regime, the contrast between the direct echo of the target and the multiple scattering background in the confocal image is therefore given by:
\begin{equation}
\label{contrast_confocal}
\mathcal{C}_I=\frac{\mathcal{I}_T}{\mathcal{I}_M} \sim  \frac{N_{\theta} N_u N_f}{N_S N_T} \frac{\sigma_T^2}{\sigma_M^2}
\end{equation}

\subsection{Target contrast provided by matrix fingerprint imaging}

To evaluate the signal-to-noise ratio provided by the matrix fingerprint imaging process, we will consider Eq.~\ref{eq P} with a fingerprint operator equal to the target matrix: $\mathbf{F}=\mathbf{R}_0$. The mean intensity $\mathcal{F} $ of the fingerprint image is then given by:
\begin{equation}
\label{mean_fingerprint}
\mathcal{F} = \beta^{-1} \left \langle  \left | \frac{1}{\Delta f}\int_{f_-}^{f_+} df \sum_{\bm{\theta}_{\textrm{in}}}^{N_{\theta}}\sum_{\mathbf{u}_\textrm{out}}^{N_u} R(\mathbf{u}_\textrm{out},\bm{\theta}_{\textrm{in}},f) R_0^*(\mathbf{u}_\textrm{out},\bm{\theta}_{\textrm{in}},f)\right |^2 \right \rangle
\end{equation}
where $\beta$ stands for the normalization factor of Eq.~\ref{eq P}.

For the multiple scattering contribution, the associated mean intensity $\mathcal{F}_M$ can be computed by considering $\mathbf{R}=\mathbf{M}$ in Eq.~\ref{mean_fingerprint}
\begin{equation}
\label{mean_fingerprint2}
\mathcal{F}_M =\beta^{-1} \left \langle \left | \frac{1}{\Delta f}\int_{f_-}^{f_+} df \sum_{\bm{\theta}_{\textrm{in}}}^{N_{\theta}}\sum_{\mathbf{u}_\textrm{out}}^{N_u} M(\mathbf{u}_\textrm{out},\bm{\theta}_{\textrm{in}},f) R_0^*(\mathbf{u}_\textrm{out},\bm{\theta}_{\textrm{in}},f)\right |^2 \right \rangle
\end{equation}
The reflection matrix coefficients of the multiple scattering contribution and of the target component are fully uncorrelated. The triple sum is therefore incoherent in Eq.~\ref{mean_fingerprint2} :
\begin{equation}
\label{mean_fingerprint3}
\mathcal{F}_M = \beta^{-1} \frac{1}{N_f} \sum_{\bm{\theta}_{\textrm{in}}}^{N_{\theta}}\sum_{\mathbf{u}_\textrm{out}}^{N_u} \left \langle |M(\mathbf{u}_\textrm{out},\bm{\theta}_{\textrm{in}},f)|^2 \right \rangle  \left \langle|R_0(\mathbf{u}_\textrm{out},\bm{\theta}_{\textrm{in}},f)|^2 \right \rangle
\end{equation}
which simplifies into:
\begin{equation}
\label{mean_fingerprint4}
\mathcal{F}_M \sim \frac{N_{\theta} N_u}{\beta N_f}  \sigma_M^2 
\end{equation}

For the target contribution, the associated fingerprint intensity $\mathcal{F}_T$ can be derived by considering $\mathbf{R}=\sigma_T \mathbf{R}_0$ in Eq.~\ref{mean_fingerprint}:
\begin{equation}
\label{mean_fingerprint5}
\mathcal{F}_T = \frac{\sigma_T^2}{\beta} \left \langle \left | \frac{1}{\Delta f}\int_{f_-}^{f_+} df \sum_{\bm{\theta}_{\textrm{in}}}^{N_{\theta}}\sum_{\mathbf{u}_\textrm{out}}^{N_u} \left |R_0(\mathbf{u}_\textrm{out},\bm{\theta}_{\textrm{in}},f) \right |^2\right |^2 \right  \rangle 
\end{equation}
Each term in the triple sum is coherent, which yields the following scaling for the target intensity $\mathcal{F}_T$:
\begin{equation}
\mathcal{F}_T \sim \frac{N_{\theta}^2 N_u^2}{\beta} \sigma_T^2
\end{equation}

The contrast between the target signal with respect to the multiple scattering background (Eq.~\ref{mean_fingerprint4}) in the likelihood image is therefore given by:
\begin{equation}
\label{contrast_final}
\mathcal{C}_\gamma=\frac{\mathcal{F}_T}{\mathcal{F}_M} \sim  N_{\theta} N_u N_f\frac{\sigma_T^2}{\sigma_M^2}
\end{equation}
Compared to the confocal image (Eq.~\ref{contrast_confocal}), the gain $G$ provided by the fingerprint operator in terms of contrast between the target and the multiple scattering fog scales as:
\begin{equation}
\label{GG}
G=\frac{\mathcal{C}_\gamma}{\mathcal{C}_I} \sim  N_S N_T
\end{equation}
$G$ corresponds to the number of spatial and temporal d.o.f exhibited by the target echo. 

\section{Precision of the localization process}
\label{Cramer}

As shown by Desailly and collegues~\cite{Desailly2015} in the context of ultrasound localization microscopy~\cite{errico_ultrafast_2015}, the transverse and axial localization precision can be expressed by means of the Cram\'{e}r-Rao bound as follows:
\begin{equation}
\label{transverse}
\delta \rho_L \sim 2\sqrt{3} \frac{\sigma_t}{f_c}  \frac{\delta \rho_D}{\sqrt{N_u N_{\theta}}}
\end{equation}
and 
\begin{equation}
\label{axial}
\delta z_L \sim  \frac{c\sigma_t}{2} \frac{1}{\sqrt{N_u N_{\theta}}}
\end{equation}
where $\sigma_t$ is the standard deviation $\sigma_t$ of echo time estimates. This quantity has been derived in a seminal paper by Quazi for a limited bandwidth ($\Delta f \ll f_c$)~\cite{Quazi1981}:
\begin{equation}
\label{Quazi}
\sigma_t=\frac{1}{2\sqrt{2}\pi} \sqrt{\frac{\sigma_M}{\sigma_T}} \frac{f_c}{\sqrt{N_f}} 
\end{equation}
Injecting Eq.~\ref{Quazi} into Eq.~\ref{transverse} leads to the following expression for $\delta \rho_L$:
\begin{equation}
\label{transverse2}
\delta \rho_L \sim \frac{1}{\pi}\sqrt{\frac{3}{2}} \sqrt{\frac{\sigma_M}{\sigma_T}}  \frac{\delta \rho_D}{ \sqrt{N_f N_u N_{\theta}}}
\end{equation}
Using Eq.~\ref{contrast_final}, the previous equation can be recast as follows:
\begin{equation}
\label{transverse3}
\delta \rho_L \sim  \frac{1}{\pi}\sqrt{\frac{3}{2}} \frac{\delta \rho_D}{\sqrt{\mathcal{C}_\gamma}}
\end{equation}
As regards to the axial resolution $\delta z_L$, Eqs.~\ref{Quazi} and \ref{contrast_final} can be injected into Eq.~\ref{axial} to obtain:
\begin{equation}
\label{axial2}
\delta z_L \sim  \frac{1}{4\sqrt{2}\pi} \frac{c f_c}{\sqrt{\mathcal{C}_\gamma}}
\end{equation}
The transverse and axial precision of the localization process are therefore inversely proportional to the target contrast. This fundamental result shows the benefit of the fingerprint operator for localization purposes.

\section{Detection threshold of potential targets}
\label{PFA}

\begin{figure*}[ht]
\centering
\includegraphics[width=10cm]{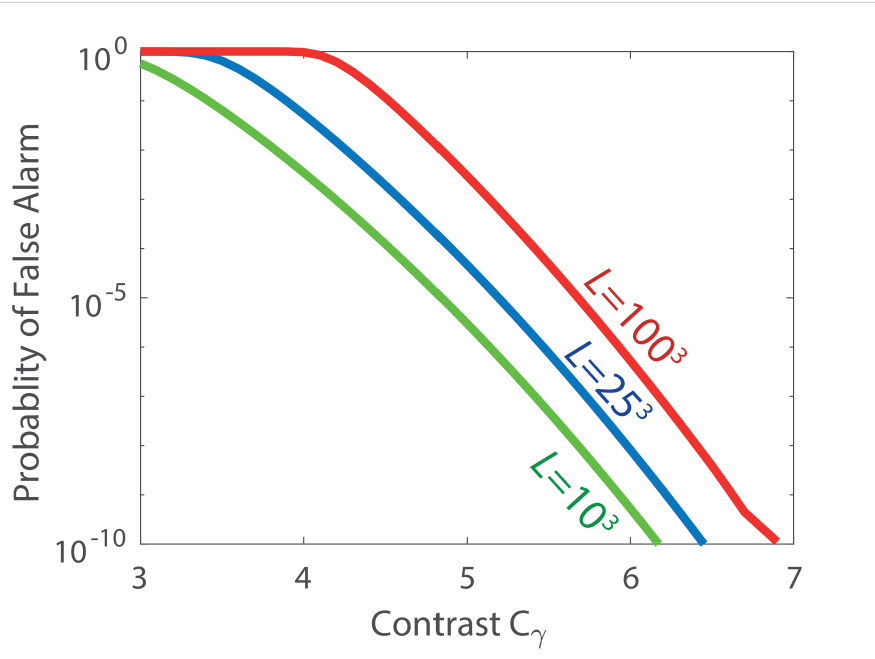}
\caption{\label{PFA2} {\textbf{Probability of false alarm.} The probability of false alarm, $1-|F_R(\gamma^2)|^L$, is displayed s a function of $\mathcal{C}_{\gamma}=(\gamma_{\textrm{max}}/\gamma_M)^2$ for different values of $L$, the number of independent pixels in the field-of-view: $L=10^3$ (green), $L=25^3$ (blue), $L=100^3$ (red).}}
\end{figure*}

\revv{Since the scattering medium is considered as one realization of a random process, assessing the reliability of the obtained images requires a statistical model for the probability density function of the $\gamma$-map in the multiple scattering regime. If the reflection matrix coefficients are independently and identically distributed random variables, the intensity $\gamma^2$ of each image pixel would obey the Rayleigh distribution in absence of any target:
\begin{equation}
\rho_R(\gamma^2)=\sqrt{\frac{2}{\pi}}\frac{\gamma^2}{\gamma_M^2} \exp \left ( -\frac{\pi}{4} \frac{\gamma^4}{ \gamma_M^4} \right )
\end{equation}
with $\gamma_M= \sqrt{\langle \gamma^2 \rangle}$, the mean value of $\gamma$ associated with the multiple scattering background. As a detection threshold based on the maximum value $\gamma_\textrm{max}$ of the $\gamma$-map is needed, the statistical distribution of $\gamma_\textrm{max}^2$ has to be known in the multiple scattering regime. As each pixel are independent from each other in the multiple scattering regime, then the distribution function $F_{\textrm{max}}$ of $\gamma_\textrm{max}^2$ is simply given by the $L^{\textrm{th}}$ power of the distribution function $F_R$ of one pixel,
\begin{equation}
\label{FR}
F_R (\gamma^2)=1-\exp \left ( -\frac{\pi}{4} \frac{\gamma^4}{ \gamma_M^4} \right )
\end{equation}
with $L$, the number of independent speckle spots in the field-of-view. The quantity $1-[F_R(\gamma_\textrm{max}^2)]^L$ is the probability that a pure multiple-scattering speckle gives rise to a likelihood index larger than the measured $\gamma_{\textrm{max}}$. Therefore, $[F_R(\gamma_\textrm{max}^2)]^L$ can be directly used as a probability that the brightest pixel is actually associated with a coherent target rather than a normal multiple-scattering speckle fluctuation. In the experiment considered in Figure 3, we have $\mathcal{C}_{\gamma}=(\gamma_{\textrm{max}}/\gamma_M)^2 \sim 6$ and $L\sim 25^3$. Applying these numerical values to Eq.~\ref{FR} yields a probability of $10^{-8}$ that the brightest spot is associated with a false alarm (see Fig.~\ref{PFA2}). Each sphere is therefore detected with an extremely high degree of confidence.}

\section{Link between the mapping of the likelihood index and a confocal beamforming process}
\label{supp_link}

{In terms of computational efficiency, the best strategy is to compute the likelihood index $\gamma$ directly in the plane wave basis. Indeed, Eq.~\ref{eq P} can then be written in terms of matrix coefficients as follows:
\begin{equation}
\gamma(\mathbf{r})=\beta^{-1} \int df \sum_{\theta_{\textrm{in}}}\sum_{\theta_{\textrm{out}}} R(\theta_{\textrm{out}},\theta_{\textrm{in}},f) F^*(\theta_{\textrm{out}},\theta_{\textrm{in}},\mathbf{r},f) 
\end{equation}
with $\beta$, a normalization coefficient. Injecting Eq.~\ref{BF} into this last equation leads to the following expression for the likelihood index
\begin{equation}
  \gamma(\mathbf{r}) = \beta^{-1} \int df {S}(\bm{\theta}_{\textrm{out}},\Delta \mathbf{r},f) R(\theta_{\textrm{out}},\theta_{\textrm{in}},f) {{F}}^* (\bm{\theta}_{\textrm{out}},\bm{\theta}_{\textrm{in}},\mathbf{r}_0,f) {S}(\bm{\theta}_{\textrm{in}},\Delta \mathbf{r},f).
\end{equation}
The last equation can be rewritten as the following matrix product:
\begin{equation}
\label{gamma_BF}
  \gamma(\mathbf{r}) = \beta^{-1} \int df  \mathbf{S}^{\dag}(\mathbf{r}-\mathbf{r}_0,f) \times \left [\mathbf{R}_{\theta \theta} (f) \circ \mathbf{F}^*_{\theta \theta}(\mathbf{r}_0,f) \right] \times  \mathbf{S}^{\dag}(\mathbf{r}-\mathbf{r}_0,f) 
\end{equation}
This last equation shows a strong analogy with a confocal beamforming process that write as follows when operated from the plane wave basis:
\begin{equation}
\label{I_BF}
  \mathcal{I}(\mathbf{r}) = \beta^{-1} \int df  \mathbf{S}^{\dag}(\mathbf{r},f) \times \mathbf{R}_{\theta \theta} (f) \circ  \times  \mathbf{S}^{\dag}(\mathbf{r},f) 
\end{equation}
The comparison between Eqs.~\ref{gamma_BF} and \ref{I_BF} shows that the spatial mapping of the likelihood index is equivalent to a confocal beamforming process applied to the Hadamard product between the reflection matrix $\mathbf{R}_{\theta \theta} (f)$ and the fingerprint operator $\mathbf{F}_{\theta \theta} (f)$. Compared to standard confocal beamforming, the computational burden of our approach only consists in the computation of this Hadamard product.} 

\clearpage

\def\bibsection{\noindent \textbf{References.}}

%

\end{document}